\pdfoutput=1
\documentclass[10pt,journal,compsoc]{IEEEtran}
\usepackage{amsmath,amsfonts}
\usepackage{algorithmic}
\usepackage{algorithm}
\usepackage{array}
\usepackage[caption=false,font=normalsize,labelfont=sf,textfont=sf]{subfig}
\usepackage{textcomp}
\usepackage{stfloats}
\usepackage{url}
\usepackage{verbatim}
\usepackage{graphicx}
\usepackage{cite}
\graphicspath{{figures/}{pictures/}{images/}{./}} 

\usepackage{microtype}                 
\PassOptionsToPackage{warn}{textcomp}  
\usepackage{textcomp}                  
\usepackage{mathptmx}                  
\usepackage{times}                     
\usepackage{cite}                      
\usepackage{tabu}                      
\usepackage{booktabs}                  

\usepackage{amssymb}
\usepackage{amsmath}
\usepackage{comment}
\usepackage[dvipsnames]{xcolor}
\usepackage{subfig}
\usepackage[colorlinks = true,
            linkcolor = blue,
            urlcolor  = blue,
            citecolor = blue,
            anchorcolor = blue]{hyperref}

\usepackage{enumitem}

\usepackage{etoolbox}
\usepackage{pdfpages}

\newcommand{\figsize}{0.2}

\begin{document}

\title{Fast Time-Varying Contiguous Cartograms\\ Using Integral Images}

\author{Vladimir~Molchanov, Hennes~Rave, and Lars~Linsen
  \thanks{All authors are with the University of Münster, Germany. \protect\\
    E-mail: \{hennes.rave\textbar molchano\textbar linsen\}@uni-muenster.de.}
}



\maketitle

\begin{abstract}
  Cartograms are a technique for visually representing geographically distributed statistical data, where values of a numerical attribute are mapped to the size of geographic regions. Contiguous cartograms preserve the adjacencies of the original regions during the mapping. To be useful, contiguous cartograms also require approximate preservation of shapes and relative positions. Due to these desirable properties, contiguous cartograms are among the most popular ones. Most methods for constructing contiguous cartograms exploit a deformation of the original map. Aiming at the preservation of geographical properties, existing approaches are often algorithmically cumbersome and computationally intensive. We propose a novel deformation technique for computing time-varying contiguous cartograms based on integral images evaluated for a series of discrete density distributions. The density textures represent the given dynamic statistical data. The iterative application of the proposed mapping smoothly transforms the domain to gradually equalize the temporal density, i.e., region areas grow or shrink following their evolutionary statistical data. Global shape preservation at each time step is controlled by a single parameter that can be interactively adjusted by the user. Our efficient GPU implementation of the proposed algorithm is significantly faster than existing state-of-the-art methods while achieving comparable quality for cartographic accuracy, shape preservation, and topological error. We investigate strategies for transitioning between adjacent time steps and discuss the parameter choice. Our approach applies to comparative cartograms’ morphing and interactive cartogram exploration.
\end{abstract}


\section{Introduction}
\label{sec:intro}


Since ancient times, geographical data visualization has been serving for documenting knowledge of the surrounding world at different spatial scales. With advances in technology, geographical maps have become more accurate, informative, and specialized. Besides locations of various objects (e.g., cities, mountains, ocean currents), geographical maps depict the objects' properties (e.g., elevation, temperature). Such data allow for a fast and direct comparison of objects of the same type. However, some statistical data, for instance, country population, is to be attributed to larger map regions rather than to certain locations. Then, visual representations of such integral statistics require techniques to encode both the statistical value and the respective region.

One common approach to convey geographically distributed integral statistical data is to use color, as it is done in choropleth maps. However, information in choropleth maps can be misinterpreted due to high discrepancy of region sizes, the so-called {area size bias}~\cite{Schiewe23}. Moreover, originally continuous data is often binned to reduce the number of used color shades and to avoid problems related to color perception~\cite{Muller79}.
In contrast to other types of thematic maps, \emph{cartograms} use area encoding for the representation of continuous numerical data. Thus, geographical regions are shown as geometrical shapes where areas are proportional to the attributed statistical values. An example of a contiguous cartogram is given in Figure~\ref{fig:teaser}, where population data in France is encoded using the approach proposed in this work.

For the exploration of the temporal evolution of geographical data or comparative analysis of several spatial statistics, multiple cartograms can be constructed for the same original map. For interactive analysis, on-the-fly generation of animations and interactive control of the cartogram's parameters are required. Efficient morphing of cartograms may dramatically improve accuracy in synoptic tasks~\cite{Duncan21}. Therefore, the following characteristics of the deformation algorithm for the construction of contiguous cartograms become crucial \cite{Nickel22}: \emph{Algorithmic complexity and scalability}, \emph{speed of computations}, and \emph{stability}. Another challenge is to ensure the \textit{coherence} of cartograms computed for consecutive time steps. The coherence is necessary for visual comparative analysis of time-varying cartograms.

In an earlier paper~\cite{Molchanov20_wscg}, we used Integral Images (InIms) to deform a single geographical map. The proposed mapping did not converge but might be used for constructing approximate cartograms. In our follow-up paper~\cite{Rave24_TVCG}, we improved the numerical scheme resulting in a convergent iterative density-equalizing transformation~(DET). We applied this transformation for the regularization of scatterplots. In our current work presented here, we adapt the DET based on InIms for computing accurate contiguous cartograms.
Global shape preservation is controlled by a single parameter, which can be interactively adjusted by the user. We perform numerical tests examining the properties and efficiency of our proposed method and evaluate our technique in comparison to the state of the art.
The proposed algorithm has low computational complexity and is therefore suitable for generating animated morphings at runtime. Thus, we apply our method for computing contiguous cartograms for dynamic statistical data. We study different approaches for ensuring the coherence of individual time steps and discuss and numerically evaluate the choices of our algorithm's parameters.

\noindent
The individual contributions of our work can be summarized by:
\begin{itemize}
  \item We introduce a novel deformation mapping approach based on integral images, enabling the transformation of original geographical maps into contiguous cartograms aligned with specified statistical data.
  \item  The algorithm is generalized to handle time-varying data, making it applicable to dynamic scenarios.
  \item  We design intuitive views and user interactions to control and adjust key parameters of the algorithm for both static and temporal datasets.
  \item  Comprehensive numerical tests are conducted to evaluate the properties and efficiency of the proposed method.
  \item  We analyze and quantify the impact of various algorithm parameters through numerical evaluations.
  \item  We compare our technique to state-of-the-art methods, highlighting its advantages and differences.
\end{itemize}

\newcommand{\figheight}{3.4cm}
\begin{figure}[!t]
  \centering
  \includegraphics[height=\figheight]{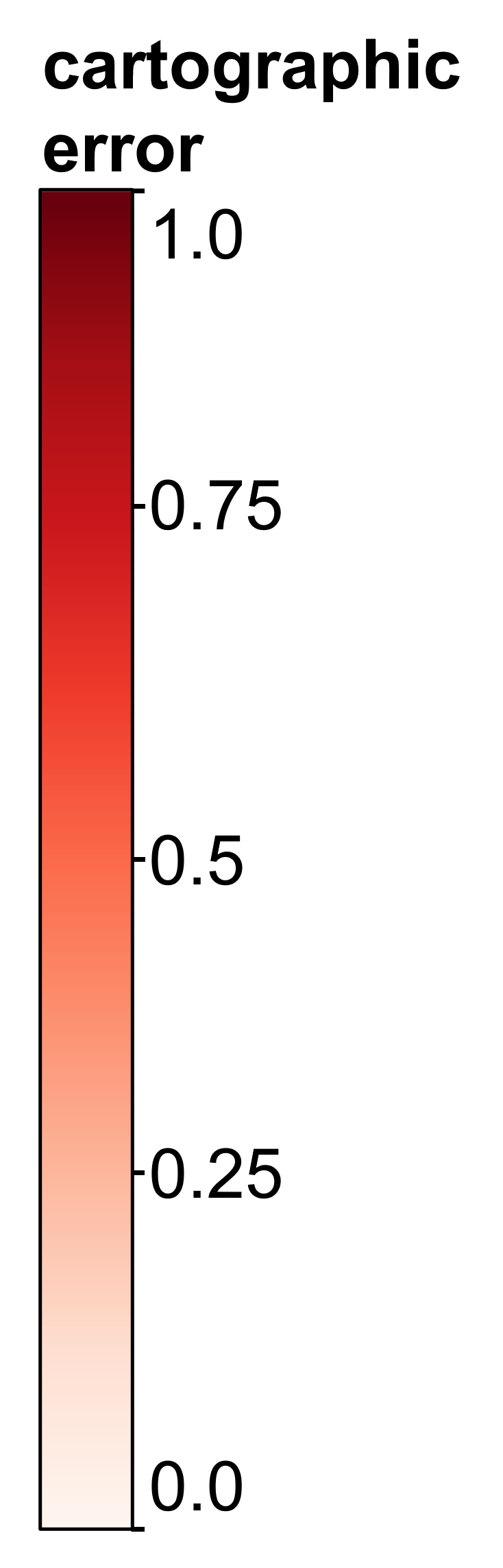}
  \subfloat[choropleth map]{\includegraphics[height=\figheight]{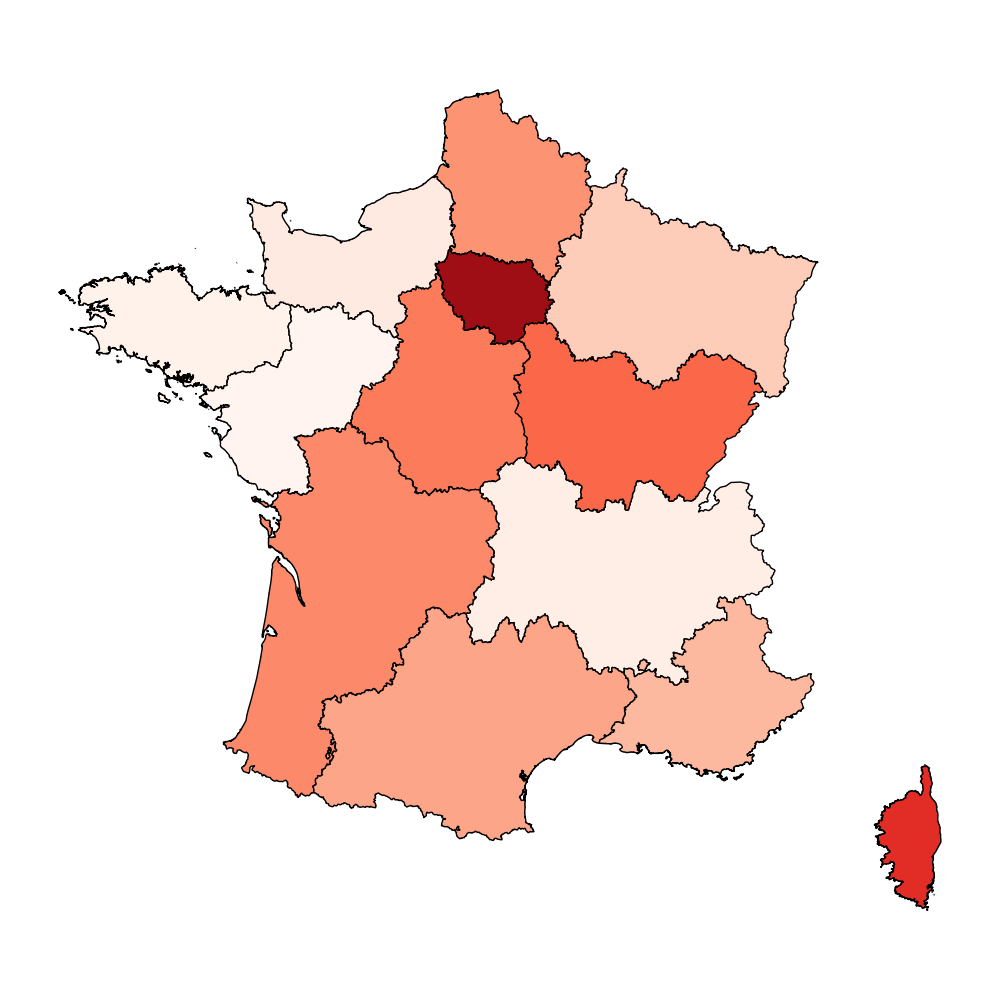} \label{fig::fr_orig}}\
  \subfloat[4 iter., 14.35 ms]{\includegraphics[height=\figheight]{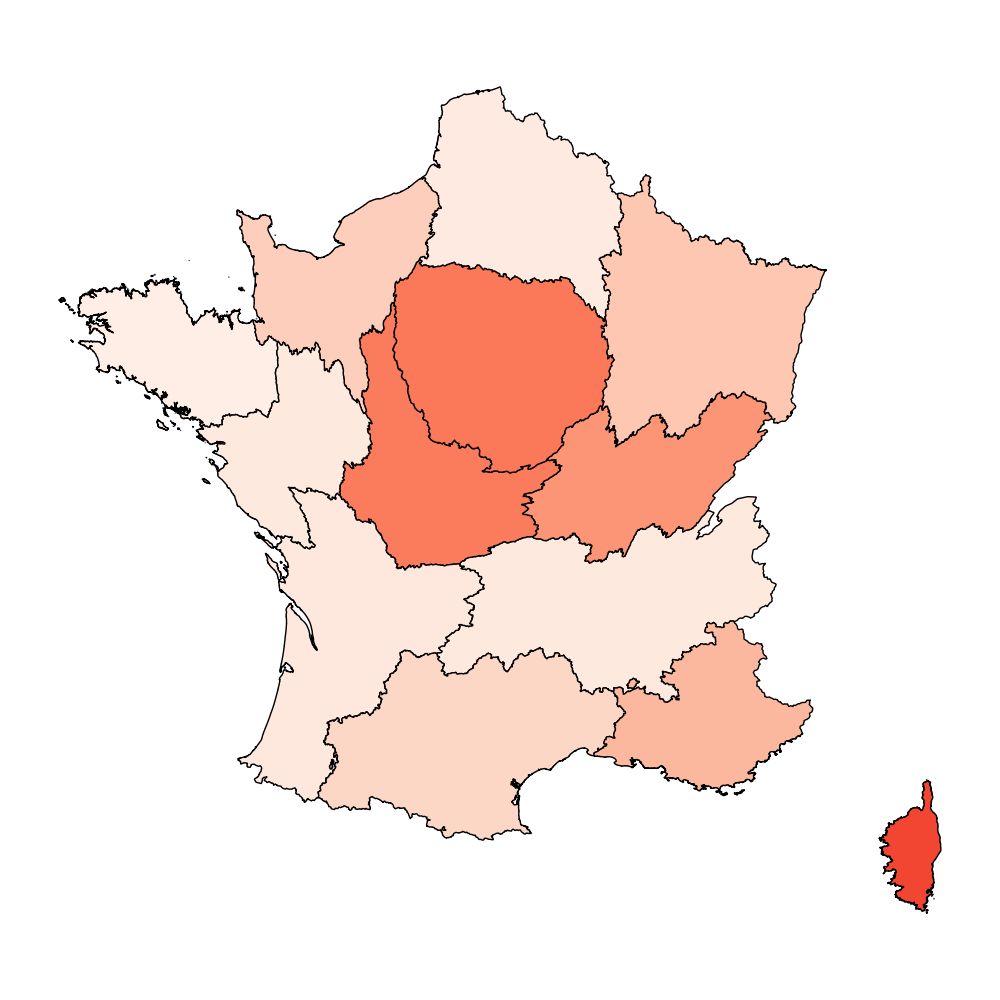} \label{fig::fr_4}}\\
  \subfloat[16 iter., 41.88 ms]{\includegraphics[height=\figheight]{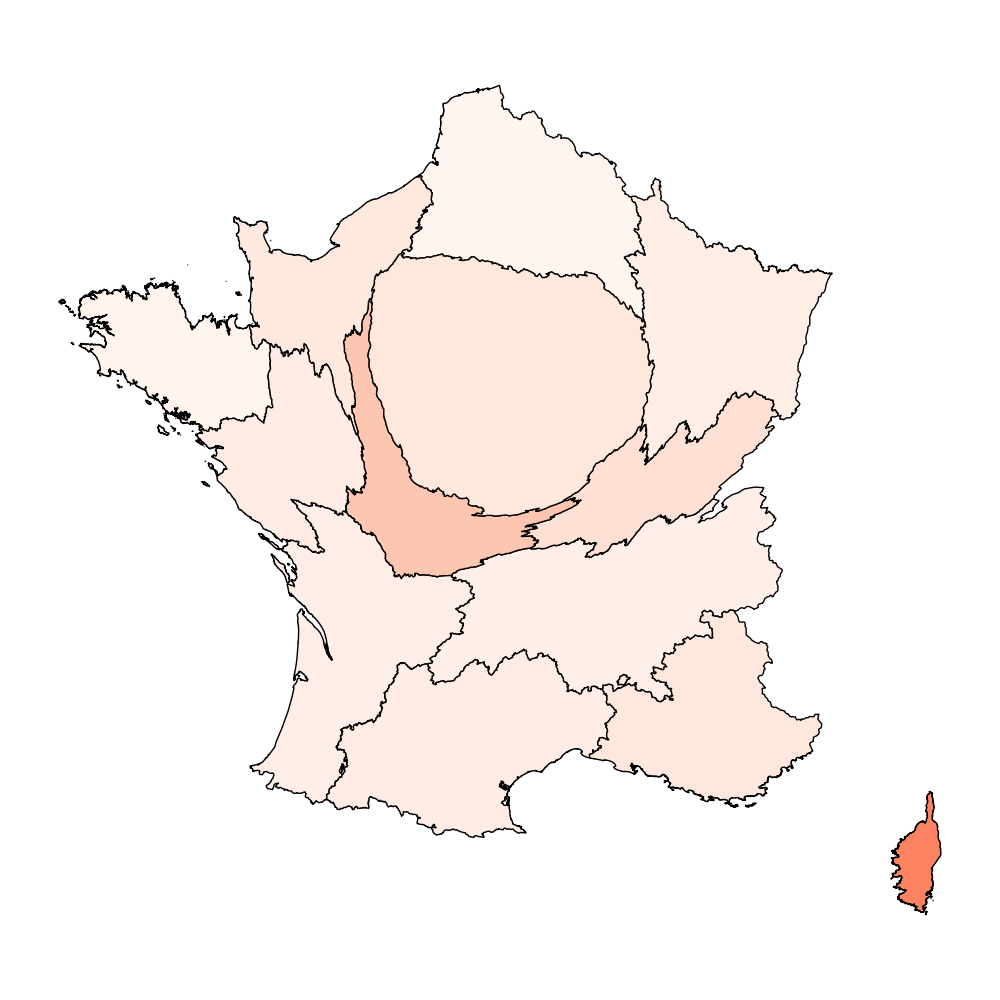} \label{fig::fr_16}}\
  \subfloat[64 iter., 123.57 ms]{\includegraphics[height=\figheight]{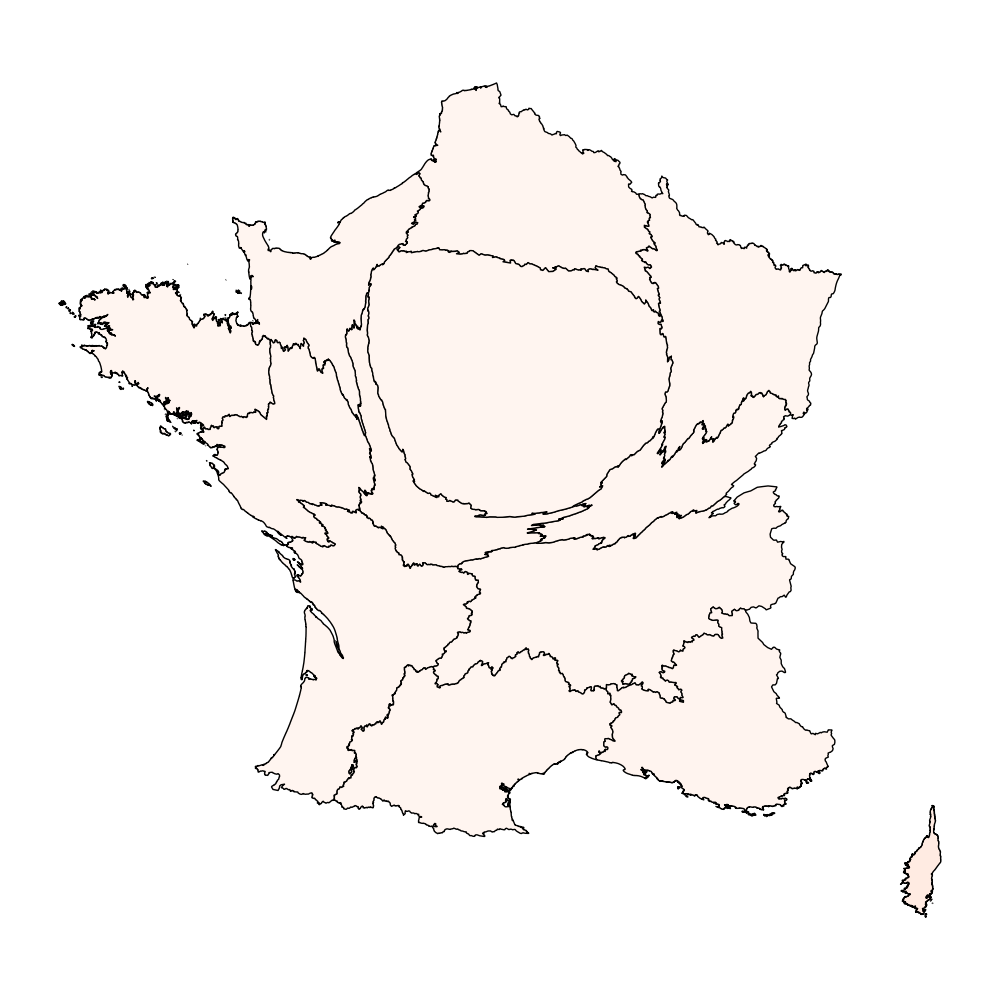} \label{fig::fr_164}}
  \caption{Iterative construction of a contiguous cartogram for population data in France~\cite{gocart}: Our approach makes regions with high population grow and regions with low population shrink, while also aiming at shape and topology preservation. Computation times allow for application in interactive analyses.}
  \label{fig:teaser}
\end{figure}

\section{Related Work} \label{sec:related_work}

Cartograms (or \emph{value-by-area maps}) encode values of a numerical variable associated with geographic regions as areas of the regions. Most of the cartograms fall into one of the following types: contiguous, non-contiguous, Dorling, Raisz, mosaic, or semi-contiguous cartograms~\cite{Kreveld07, Gastner21}. Transformation of the original map is usually the main idea in constructing \emph{contiguous} cartograms, which have been the most commonly used cartograms for years~\cite{Nusrat16}.
Nusrat and Kobourov~\cite{Nusrat16} surveyed cartogram research and developed a taxonomy of tasks. Recently, good practices for contiguous cartograms were summarized by Tingsheng et al.~\cite{Tingsheng20}. The authors emphasized that cartograms should be used to represent additive statistics with interpretable totals. Showing the original map which uses the same color coding as the cartogram helps to understand the encoded data. Missing data should be addressed and a legend should present the unit area magnitude. Finally, interactive tools are useful for electronically delivered cartograms.

Approaches for the construction of contiguous cartograms vary significantly, ranging from image-based operations~\cite{Sagar14, Molchanov20_wscg} and computational geometry transformations~\cite{Tobler86, GuseinZade93, Keim05} to artificial intelligence~\cite{Henriques09} and algorithms motivated by physical process simulations~\cite{Gastner04, Sun13, Sun13Optimized, Gastner18}.

\textbf{Pseudo-Cartograms} are distortions of the original geographic maps based on given statistical data, which are limited in their application. For instance, Tobler~\cite{Tobler86} derived an analytic formula for the optimal DET in the class of separable functions. The resulting map is referred to as \emph{pseudo-cartogram}, since the cartographic error vanishes only for separable density distributions. Nevertheless, Tobler suggested using pseudo-cartograms as a starting point for more sophisticated cartogram construction algorithms to improve their performance in terms of computation times.
\emph{RadialScale} and \emph{AngularScale} approaches proposed by Bak et al.~\cite{Bak09} distort a map in radial and angular directions, thus, constructing a separable mapping in polar coordinates.

Keim et al. developed several algorithms, including \emph{HistoScale} for faster computation of pseudo-cartograms~\cite{Keim03}, \emph{CartoDraw} for better preservation of the shapes of the regions using a Fourier transformation of the boundary curvatures~\cite{Keim04}, its modification \emph{Medial-Axis-Based} approach~\cite{Keim05}, and \emph{VisualPoints}, which outperforms \emph{CartoDraw} with respect to computation time and area error~\cite{Keim03_vgi}.

Sagar~\cite{Sagar14} proposed a region-growing morphology-based algorithm. The set of regions is replaced by a set of their centroids on a raster image. Then, influence zones around each centroid are computed iteratively via dilation operations. The speed of growth of each zone is determined by the statistical value in the respective region. The method is computationally expensive~\cite{Sagar14} and has a non-trivial cartographic error~\cite{Nusrat16}. In our earlier paper~\cite{Molchanov20_wscg}, we proposed to use InIms of a density texture for map distortion. When the density function reflects the distribution of a given regional statistical data, the distorted map can serve as a pseudo-cartogram. The method is computationally fast and suitable for interactive applications. The proposed transformation is density \textit{decreasing}, but does not generally lead to vanishing cartographic error.

\textbf{Contiguous Cartograms} are usually constructed through an iterative process of consecutive map deformations gradually decreasing the cartographic error. For example, the concept of rubber-sheet methods is based on simulating a system of springs attached to the vertices at the region boundaries. The vertices are then gradually moved in a force field generated from the statistics distribution. As the areas of the transformed regions become close to the desired sizes, a minimal energy state is reached and the force field magnitude becomes negligible. The original rubber-sheet algorithm proposed by Dougenik~\cite{Dougenik85} cannot guarantee topology preservation and non-overlapping regions. Sun~\cite{Sun13Optimized} introduced a new mathematically deduced parameter -- the global elasticity coefficient -- to improve topological integrity and optimize the convergence rate of the classical algorithm. The \emph{Carto3F} method by Sun~\cite{Sun13} eliminates topological errors and computationally outperforms the former algorithm by using a quadtree structure and parallel computations.

The Gastner and Newman diffusion method~\cite{Gastner04} remains one of the most popular algorithms for generating contiguous cartograms~\cite{Nusrat16}. The method simulates a density-equalizing flow directed from high-density regions toward low-density areas. The resulting cartograms are accurate and topologically correct. However, the shapes of the regions may be deformed significantly, which affects the cartogram readability negatively. The flow-based algorithm by Gastner et al.~\cite{Gastner18} is a recent example of physics-inspired cartogram methods. Instead of the heat equation used in~\cite{Gastner04}, the authors simulate the density-equalizing process using the continuity equation. The desired map transformation is obtained as a finite-time integral of a velocity field, which linearly equalizes the given density distribution towards its mean.

The comparison of algorithms for computing cartograms is ambiguous, since no standard set of quantitative metrics exists and different approaches significantly vary in computational efficiency, ease of implementation, and performance~\cite{Henriques09}. Alam et al.~\cite{Alam15} distinguished seven quantitative measures that capture different properties of cartograms: Average and maximum cartographic errors, adjacency and angular orientation errors, Hamming distance, average aspect ratio, and polygonal complexity. The effectiveness of cartograms of different types was examined by Sun and Li~\cite{Sun10} in a user study. In particular, the authors reported that among the contiguous cartograms, the diffusion cartogram by Gastner and Newman~\cite{Gastner04} performs better than the rubber sheet cartogram by Dougenik~\cite{Dougenik85}. In the user study performed by Nusrat et al.~\cite{Nusrat18}, the participants were asked to evaluate cartograms' helpfulness of visualization, readability, and appearance. The results demonstrated a clear preference for contiguous and Dorling cartograms~\cite{Dorling96}.

In summary, we conclude that the diffusion cartogram approach by Gastner and Newman~\cite{Gastner04} is a traditional approach that achieves desirable results and, thus, has been widely used. The more recent flow-based algorithm by Gastner et al.~\cite{Gastner18} achieves a major speed-up and can be considered the state of the art for contiguous cartograms. We compare our approach against those approaches in Section~\ref{sec:results}.

\textbf{DETs} are core for most cartogram construction methods. Gusein-Zade and Tikhunov~\cite{GuseinZade93} proposed an iterative approach to computing a geometric transformation whose Jacobian is equal to the given density function. They mathematically demonstrated that the solution to the general problem of the contiguous cartogram construction is not unique.

Our proposed algorithm adopts and adapts the \textbf{regularization} algorithm developed by Rave et al.~\cite{Rave24_TVCG} for data regularization in scatterplots. The spatial transformation depends analytically on values of the \emph{integral images} (InIms) computed from the rasterized density distribution. InIms were originally proposed by Crow~\cite{Crow84} and Viola and Jones~\cite{Viola02} for object detection in image analysis. Computation of InIms at arbitrary angles was studied by Chin et al.~\cite{Chin08}. Ehsan et al.~\cite{Ehsan15} addressed the problem of efficient parallel computation of InIms. Singhal et al.~\cite{Singhal12} discussed the implementation of the computational algorithms on the GPU.

Recent density-equalizing regularization algorithms for occlusion reduction in scatterplots include the space-filling curves approach by Cutura et al.~\cite{Cutura21, Cutura22}, pixel maps by Raidou et al.~\cite{Raidou19}, the distance grid method by Hilasaca et al.~\cite{Hilasaca23}, and the limited distortion with optimized node size approach by Giovannangeli et al.~\cite{Giovannangeli24}. Often, these methods have high algorithmic complexity due to the need for locally resolving collisions of samples and non-linear optimization procedures. InIms-based regularization proposed by Rave et al.~\cite{Rave24_TVCG} demonstrates better scalability, lower execution times, and reliable topology preservation properties. Therefore, we adapt and examine their approach for the fast construction of contiguous cartograms.

\textbf{Interactivity} plays an important role in modern exploratory visualization systems. However, Duncan et al.~\cite{Duncan21} argued that comprehension of computer-aided cartograms would not benefit from zooming. On the other hand, smooth morphing between cartograms representing different statistics~\cite{Reilly04}, animating cartograms for temporal data~\cite{Demoraes21}, or linked brushing on the original map and the corresponding cartogram~\cite{Dykes97, Dykes98} may support the user in exploratory tasks. However, interactivity requires high efficiency of the cartogram construction algorithms regarding scalability, complexity, and performance. Thus, not all existing approaches qualify to be used in interactive systems. Due to its high computational efficiency, our method is designed for interactive applications.

\section{Background} \label{sec:background}

Rave et al.~\cite{Rave24_TVCG} proposed an algorithm for regularizing sample distribution in scatterplots. The regularization procedure iteratively deforms the scatterplot domain such that the overall distribution of samples becomes nearly uniform. The goal of the algorithm is to mitigate occlusion in scatterplots. The authors implemented the proposed technique on the GPU and reported interactive frame rates even for large datasets. We adapt the proposed DET for fast construction of contiguous cartograms. We use geostatistical data for computation of the density texture following the pseudo-cartogram construction by Molchanov and Linsen~\cite{Molchanov20_wscg}. This section provides all necessary technical details of the original method, in particular, it introduces InIms (Section~\ref{sec:inims}), which are in the core of the technique, and the domain transformation formula (Section~\ref{sec:meshmapping}). For the sake of readability and consistency, we preserve the notations used by Rave et al.~\cite{Rave24_TVCG}. Furthermore, in this section, we also provide the quality metrics formulae, which we use for numerical comparison of the resulting method with state-of-the-art algorithms (Section~\ref{sec:quality}).

\subsection{Integral Images}\label{sec:inims}

\begin{figure}[!bt]
  \centering
  \includegraphics[width=\linewidth]{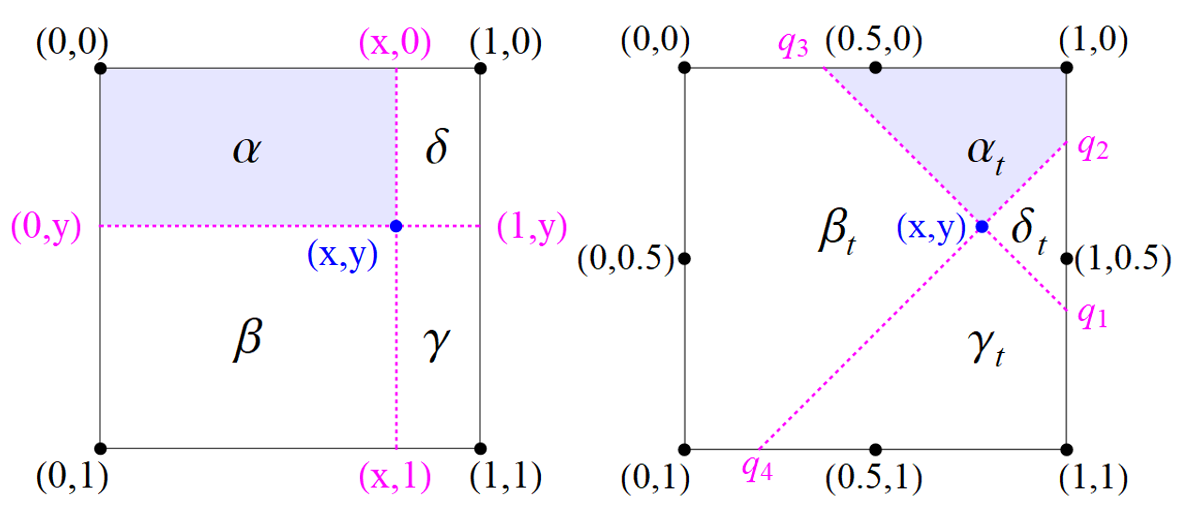}
  \caption{Sketch of InIm computations (reproduction from Molchanov and Linsen~\cite{Molchanov20_wscg}):~Areas of summation of InIms (left) and tilted InIms (right) for location $(x,y)$. The eight anchor points are shown in magenta. The integration domains partition the texture space into four disjoint regions each. The attraction force associated with anchor $q_1$ is $\alpha$,  InIm value $\beta$ corresponds to anchor $q_2$, etc., while large values of $\alpha_t$ result in a deformation vector pointing towards anchor $(x,\,1)$, large tilted InIm values $\beta_t$ increase the deformation vector component in the direction of $(1,\,y)$, etc.}\label{fig::ii_both}
\end{figure}

InIm $\alpha(i,\,j)$ is a table of precomputed sums of pixels of a given real-valued texture $d(i,\,j)$ over rectangular regions, more precisely:
\begin{equation*}\label{alpha}
  \alpha(i,\,j) = \sum\limits_{i'\leq i} \sum\limits_{j'\leq j} d(i',\,j').
\end{equation*}
Texture $\alpha$ corresponds to the top-left rectangular summation areas relative to a given location. Analogously, summing up the pixels values over bottom-left, bottom-right, and top-right rectangles results in three InIms denoted as $\beta$, $\gamma$, and $\delta$, see Figure~\ref{fig::ii_both}~(left). The respective formulae can be written as
\begin{eqnarray*}
  &\beta(i,\,j) = \sum\limits_{i'\leq i} \sum\limits_{j'>j} d(i',\,j'),\qquad
  \gamma(i,\,j) = \sum\limits_{i'> i} \sum\limits_{j'> j} d(i',\,j'),&\\
  &\delta(i,\,j) = \sum\limits_{i'> i} \sum\limits_{j'\leq j} d(i',\,j').&
\end{eqnarray*}
InIms can also be defined for rectangular summation areas rotated by $45^\circ$, which are called \emph{tilted InIms} $\alpha_t$, $\beta_t$, $\gamma_t$, and $\delta_t$, see Figure~\ref{fig::ii_both}~(right).

Obviously, the sum of the InIms (and tilted InIms) is constant for any indices $i$ and $j$, i.e.,
\begin{equation}\label{sum}
  \alpha + \beta + \gamma + \delta = \alpha_t + \beta_t + \gamma_t + \delta_t = \sum\limits_{i',\,j'} d(i',\,j') \equiv C.
\end{equation}
Throughout the paper, we assume textures of size $2^k\times 2^k$. We use integer pixel coordinates $(i, j)$ and real-valued normalized coordinates \mbox{$(x,\,y)=2^{-k}\ (i,\,j)$} interchangeably.

\subsection{Density-Equalizing Transformation}
\label{sec:meshmapping}

The eight InIms introduced in the previous section provide globally scoped information about the density distribution relative to the selected pixel $(i,\,j)$. Based on this information, it is possible to define a spatial transformation that decreases density differences across the whole domain. Repeated application of such global transformation gradually equalizes the density until reaching a stable state, i.e., a nearly constant function. Here, the density distribution should be updated after each iteration to account for the implied distortion.

In order to define a map transformation, Molchanov and Linsen~\cite{Molchanov20_wscg} introduced anchor points $q_l(x,\,y)$, $l=1,\ldots,4$, which are the intersection points with borders of the domain of lines parallel to the domain diagonals and crossing at $(x,\,y)$, see Figure~\ref{fig::ii_both}~(right):
\begin{align*}
  \mathrm{if}\ y<x:       & \quad q_1(x,\,y)=(1,\,1+y-x); & q_3(x,\,y)=(x-y,\,0);   \\
  \mathrm{if}\ y\geq x:   & \quad q_1(x,\,y)=(1-y+x,\,1); & q_3(x,\,y)=(0,\,y-x);   \\
  \mathrm{if}\ x+y<1:     & \quad q_2(x,\,y)=(x+y,\,0);   & q_4(x,\,y)=(0,\,x+y);   \\
  \mathrm{if}\ x+y\geq 1: & \quad q_2(x,\,y)=(1,\,x+y-1); & q_4(x,\,y)=(x+y-1,\,1).
\end{align*}
Note that the $y$-axis is oriented downward. Then, the transformation computed by
\begin{align}\label{map}
  t(x,\,y;\,d) = \dfrac{1}{2\,C}\big( & \alpha\cdot q_1(x,\,y) + \beta\cdot q_2(x,\,y)\,+                                                           \\
                                      & \gamma\cdot q_3(x,\,y) + \delta\cdot q_4(x,\,y)\,+  \nonumber                                               \\
                                      & \alpha_t\cdot(x,\,1) + \beta_t\cdot(1,\,y) + \gamma_t\cdot (x,\,0) + \delta_t\cdot (0,\,y)\ \big) \nonumber
\end{align}
can be interpreted as a linear combination of eight anchor points at the domain border. For example, when the current density distribution is imbalanced and exhibits high values in the upper-left corner of the domain, InIms $\alpha$ computed for pixels in the central and lower-right part of the domain become large. Correspondingly, the weight of anchor point $q_1$ increases and those pixels will be shifted towards $q_1$. Generally, InIms depend on the position of $(x,y)$, resulting in a non-linear globally smooth transformation. Note that due to Equations~\eqref{sum}, the coefficients of the linear combination in mapping~\eqref{map} sum up to unity.

Another consideration is that globally constant density $d_0$ has to be a fixed point of any DET, i.e., when the current density distribution becomes nearly constant, the transformation should converge to the identity mapping. Note that map~\eqref{map} does not satisfy this requirement. Therefore, Rave et al.~\cite{Rave24_TVCG} constructed the transformation
\begin{equation}\label{map_res}
  (x,\,y) \mapsto (x,\,y) + t(x,\,y;\,d) - t(x,\,y;\,d_0).
\end{equation}
Here, the distortion mapping of the constant density $t(x,\,y;\,d_0)$ is subtracted, so that when $d\to d_0$, the transformation becomes the identity mapping $(x,\,y) \mapsto (x,\,y)$.

Rave et al.~\cite{Rave24_TVCG} applied DET~\eqref{map_res} to regularize scatterplots for reducing visual clutter. The authors also developed an optimized GPU implementation of the algorithm achieving interactive rates for relatively large datasets and texture sizes. In our paper, we demonstrate that accurate and fast construction of contiguous cartograms is possible using mapping~\eqref{map_res}. Our work is focused on a proper definition of the density texture, the evolution of the subregions' boundaries throughout the iterative process, quality evaluation of resulting contiguous cartograms, performing numerical tests, comparing the results with the state-of-the-art, and developing algorithms for time-varying geostatistical data.

\subsection{Quality Measures}
\label{sec:quality}

The construction of contiguous cartograms is a trade-off between the accurate encoding of statistical data in the deformed regions' areas and preserving a faithful representation of the original map~\cite{Alam15}. According to Tobler~\cite{Tobler04}, quantitative measures for evaluating cartogram quality can be assigned to three categories. First, statistical distortion can be characterized by average and maximum \emph{cartographic errors}~\cite{Alam15, Nusrat16}
\begin{eqnarray*}
  \epsilon = \dfrac{1}{|V|} \sum\limits_{v\in V} \dfrac{|o(v) - w(v)|}{\max\{o(v),\  w(v)\}}, \label{avr_error}\\
  \xi = \max\limits_{v\in V} \dfrac{|o(v) - w(v)|}{\max\{o(v),\ w(v)\}},\label{max_error}
\end{eqnarray*}
where $V$ is a set of regions $v$ with desired areas $w(v)$ and $o(v)$ is the actual area of region $v$ in the cartogram. We apply the shoelace formula for computing areas of the polygonal regions.

The second group of measures captures the \emph{geographic distortion} and includes topology distortion, shape distortion, and relative position error. Topology distortion is defined by
\begin{equation*}\label{top_error}
  \tau = 1 - \dfrac{E_c \cap E_m}{E_c \cup E_m},
\end{equation*}
where $E_c$ and $E_m$ are the adjacencies between countries in the cartogram and the map, respectively~\cite{Heilmann04, Alam15}. I.e., $E_c$ and $E_m$ denote sets of pairs of regions $(u,\,v)$, which are neighbors in the cartogram and in the original map, correspondingly. The shape distortion is computed using the Hamming distance~\cite{Skienna08, Meulemans10}. Two polygons $P_m$ and $P_c$ representing the same region in the original map and the cartogram are superimposed. Then, the normalized symmetric difference area
\begin{equation*}
  \dfrac{||P_m \triangle P_c||}{||P_m \cup P_c||} = \dfrac{||(P_m \setminus P_c) \cup (P_c \setminus P_m)||}{||P_m \cup P_c||}
\end{equation*}
is computed. This value is minimized by considering all possible relative translations by full pixels of polygons $P_m$ and $P_c$~\cite{Alam15}. Finally, the relative position error is computed based on the positions of the regions' barycenters before and after the cartogram transformation. We refer the reader to Equation~(4) in~\cite{Heilmann04}:
\begin{equation*}
  R = \dfrac{1}{\pi\cdot|V|\cdot(|V|-1)} \sum\limits_{u,v\in V, u\neq v} \left|\mathrm{angle}\left(b_c(u,v),\ b_m(u,v)\right)\right|,
\end{equation*}
where $b_c(u,v)$ is the vector between barycenters of regions $u$ and $v$ in the cartogram, $b_m(u,v)$ is the analogous vector in the original map. The function \emph{angle} computes the angle between two vectors in radians from interval $[-\pi,\ \pi)$.

The third category of quality measures are \emph{numerical properties} of the cartogram algorithm including its computational complexity, execution time, and stability. The latter means that small perturbations of statistical data shall lead to minor changes in the resulting cartogram~\cite{Misue95, Nickel22}.
We apply the measures from all three categories to evaluate the proposed cartogram construction method and to compare it to state-of-the-art approaches in Section~\ref{sec:results}.

\section{Time-Varying Contiguous Cartograms} \label{sec:method}

In this section, we propose an approach for constructing time-varying contiguous cartograms for temporal geostatistical data. We start with a method for computing a single cartogram combining properly defined density texture $d$ used for pseudo-cartograms by Molchanov and Linsen~\cite{Molchanov20_wscg} and DET~\eqref{map_res} proposed by Rave et al.~\cite{Rave24_TVCG}. Our method is iterative and converging. The crucial part of the algorithm is the \textit{background density value} (BDV) assigned to the region in the visualization domain, for which no statistical data are available. The resulting contiguous cartogram can be evaluated using quality measures from Section~\ref{sec:quality}. Then, we present and compare different strategies for coherently evolving the cartogram according to the dynamic data. We show that depending on the analysis task, the user may prefer one strategy over the other. Finally, we introduce interactive tools for task-specific steering of the visualization.

\subsection{Construction of Static Cartograms} \label{sec:single}

The input data for cartogram construction usually consists of a set of polygonal contours representing the regions' boundaries and statistical values assigned to each region. These values are assumed to be integrals of some density distribution over respective regions and to have an interpretable total~\cite{Tingsheng20}. While density may also encode data variations on ﬁner spatial scale, in the simplest case, density $d$ is a constant function within each region~\cite{Hennig13, Gastner18}. These constant density values are computed by dividing the integral statistical values by the area of the respective regions. However, the regions of interest usually do not cover the entire rectangular visual domain. Thus, a background region with no associated statistical value may appear as a region surrounding the main geographical map. The choice of the optimal density value for this region is discussed below and in Section~\ref{sec:results}.

We compute the discrete density distribution for given geographical and statistical data for a desired spatial resolution. Then, mapping~\eqref{map_res} defines a per-pixel displacement field that expands regions with density $d$ exceeding the mean density and shrinks regions with lower density. To determine the shapes of distorted regions, we compute displacements at the vertices of the regions' boundaries using bilinear interpolation and advect the vertices accordingly. For the next iteration, one needs to recompute the density distribution accounting for changes in the regions' areas, while the statistical data values remain the same.

The density value of the background region may significantly affect the resulting cartogram quality. When chosen too low, the outer border of the map is pushed to the boundary of the visual domain, which results in a highly distorted cartogram. High-density values for the background, instead, ensure the preservation of the global maps' border shape, improving the cartogram's readability. However, excessive stiffness may result in high cartographic errors. Practically, we set the default BDV equal to the mean density value $d_0$ weighted by region area and allow the user to control this parameter if necessary. We conduct several numerical tests evaluating different quality measures in dependence on the background density in Section~\ref{sec:results}.

In summary, the proposed iterative algorithm for constructing contiguous cartograms consists of the following steps:
\begin{enumerate}
  \item[0.] As input to the algorithm, assume a geographical map with a set of regions $V$ described by polygons (including the background region), a statistical value $s(v)$ for each $v\in V$, and a given desired resolution $R$ of the visual domain.
  \item For the given desired resolution $R$ of the visual domain, pre-compute $t(x,\,y;\,d_0)$ with mean density $d_0$. This operation needs to be performed only once.
  \item For each map region $v\in V$ represented by its polygonal boundary, compute its current area $o(v)$ by counting the pixels that belong to the region.
  \item Define a piecewise constant density function $d$ by dividing the given statistical value $s(v)$ of each region $v\in V$ by the area $o(v)$ of the region. Choose the BDV so that the mean density in the whole visualization domain equals to $d_0$.  Store this rasterized density distribution in a scalar texture of the desired resolution $R$.
  \item For the density texture, compute all InIms and tilted InIms according to the formulae in Section~\ref{sec:inims}.
  \item Compute the density-equalizing displacement field $t$ by applying Equation~\eqref{map_res}.
  \item Transform vertices of the polygonal boundaries of the regions according to the distortions bilinearly interpolated from displacement field $t$.
  \item If a stationary state is reached or a maximum number of iterations has been performed, stop the calculations. Otherwise, repeat the deformation procedure from Step 2.
  \item Render the resulting cartogram of resolution $R$ on screen.
\end{enumerate}

An efficient implementation of the InIm computation on the GPU is presented in detail by Rave et al.~\cite{Rave24_TVCG} with a sample code available at \href{https://github.com/hennesrave/regularization}{\color{blue}{https://github.com/hennesrave/regularization}}. When not otherwise specified, we use textures of size \mbox{$1024\times 1024$}. Original geographical maps are isotropically scaled so that their largest dimension (horizontal or vertical) fits interval [0.05,\,0.95] of the texture space. Then, the maps are centered along the other dimension. We stop progressive deformations when the maximal cartographic error drops below 0.01 or does not decrease within the last 32 iterations, or when the maximal number of iterations of 512 is reached.

\subsection{Coherence of Dynamic Cartograms}

The proposed method relies on InIms computed for a rasterized density texture. Image-based operations can be highly efficient when using modern GPUs. Thus, our approach is designed for application scenarios where speed is a crucial factor. For instance, time-varying geostatistical data requires generating a series of cartograms presented to the user as an animation.

It is common practice to compute contiguous cartograms by transforming the original geographical map. The rationale for this procedure is obvious: The resulting cartogram should be as similar as possible to the undistorted map to preserve topological relations between the regions and therefore remain recognizable and easy to read by the user. However, the analysis of time-varying data usually focuses on comparing states of consequent time steps, which requires coherence of corresponding visualizations. When different cartograms are computed from the original map independently from each other, their coherence is not guaranteed. In the following, we refer to the approach as \textit{direct}, when all cartograms are constructed by deforming the original map.

An alternative approach is to compute each consecutive cartogram by reusing the cartogram from the previous time step with updated statistical data. We refer to this approach for the cartogram generation as \textit{cumulative}. Given that the geospatial data changes gradually over time, one may expect that the resulting cartograms morph smoothly, allowing the user to track the changes easily. Additionally, the computation time should decrease especially for time intervals with slow data dynamics. A negative side effect of this approach might be an accumulation of topological distortions, making the cartogram less recognizable as the simulation time progresses. In Section~\ref{sec:stability}, we perform a series of numerical tests to evaluate both approaches described above.

It is possible to imagine short-period dynamic geostatistical data, for which the global integral $M_i$ remains constant and only regional statistics vary in time. Here, subscript $i$ designates the integral over the map region excluding background, for which no statistical data are given. For instance, hourly commuting of workers from residential to industrial areas in an isolated community. Usually, however, integral $M_i$ (e.g., total population or annual GDP of a country, number of COVID-19 cases in the world) changes over time together with its regional distribution. In Section~\ref{sec:background}, we show that the optimal BDV depends on $M_i$ for a static cartogram. Here, optimality means the highest values of quality measures, foremost the preservation of the original regions' shapes. Then, for time-varying geostatistical data, the background density should depend on dynamic $M_i(t)$ if the shape preservation is important.

In the numerical tests presented in Section~\ref{sec:integral}, we show that adapting the BDV to the current global integral has a few side effects. Most importantly, changes in the total area of the map over time do not reflect dynamics of $M_i(t)$. So, there is a trade-off between coherence and visual comparison of evolving cartograms.
An alternative approach is to properly adjust the constant BDV (or, equivalently, the total mass of the background region $M_b$) for all time steps. Our goal is to reflect the dynamics of $M_i(t)$ via expansion or shrinkage of the map. However, if the map's global integral $M_i(t)$ exhibits dramatic growth, the size of the visualization domain may be insufficient to accommodate the whole cartogram. Then, proper scaling of the map is required. In Section~\ref{sec:integral} and in the accompanying video, we present our numerical tests and results.

\subsection{Interactive Steering of Cartogram Construction}\label{sec:steering}

We address the issues of choosing the constant BDV and the proper scaling of the map by introducing interactive tools and coordinated views in our visualization. First, we propose showing the graph of $M_i(t)$ coordinated with the dynamic cartogram visualization. The user may observe the temporal advancement of the computations and track the dynamics of the cumulative statistics $M_i(t)$, which is only possible by using this widget when the cartogram area remains constant. Second, the user can select time interval $[t_0,\,t_1]$ for a more focused analysis. Based on the selected interval, the extreme values of $M_i(t)$
$$
  M_{i,\mathrm{min}}=\min\limits_{t_0\leq t\leq t_1} M_i(t),\qquad\qquad M_{i,\mathrm{max}}=\max\limits_{t_0\leq t\leq t_1} M_i(t)
$$
can be found. Then, the user may specify the values $0 < A_{i,\mathrm{min}} < A_{i,\mathrm{max}} < 1$, denoting the area fractions of the visual domain, which should be occupied by the map (excluding the background region) at corresponding time moments. In practice, we recommend reserving about 5\%--10\% of the domain for the background, i.e., $A_{i,\mathrm{max}}\approx0.9$, and $A_{i,\mathrm{min}}\in[0.1,\, 0.6]$. Note that the regions' areas are proportional to the integral mass contained in these regions when density is globally equalized. In particular, the area fraction $A_i(t)$ of the cartogram map at moment $t$ should be equal to the fraction of mass $M_i(t)$ relative to the total mass in the visualization domain $M_i(t) + M_b(t)$. Then, the integral over the background region $M_b(t)$ can be computed for each $t$ from the following equations:
\begin{eqnarray}\label{eq:Mb}
  &&A_i(t):=\dfrac{M_i(t)}{M_i(t) + M_b(t)} = (1-w)\cdot A_{i,\mathrm{min}} + w\cdot A_{i,\mathrm{max}},\\
  &&w = (M_i(t) - M_{i,\mathrm{min}}) / (M_{i,\mathrm{max}}-M_{i,\mathrm{min}}).\nonumber
\end{eqnarray}
The first equation expresses the area fraction at time $t$ as a linear combination of fractions corresponding to the extreme values of the map integrals. The second equation defines the interpolation weight $w$. Correspondingly, the constant BDV is computed by dividing $M_b(t)$ by the number of background pixels in the visualization domain.

\section{Results and Discussion} \label{sec:results}

The diffusion method by Gastner and Newman~\cite{Gastner04} has been one of the most popular techniques to create cartograms over the last years~\cite{Nusrat16}. Recently, a new flow-based method by Gastner et al.~\cite{Gastner18} appeared, and the C code of the algorithm suitable for parallel computation is available at \href{https://github.com/mgastner/cartogram-cpp}{https://github.com/mgastner/cartogram-cpp}. The flow-based method exhibits a significant speed-up and still retains the advantages of previous techniques. Therefore, it can be considered the state-of-the-art technique for constructing contiguous cartograms. We compare the proposed approach with both diffusion and flow-based methods in terms of various quality measures (Section~\ref{sec:soa}) and computational efficiency (Section~\ref{sec:performance}). Finally, we perform an analysis of the influence of scaling (Section~\ref{sec:scaling}) and the BDV (Section~\ref{sec:backvalue}) on the cartogram quality, and discuss the stability of computations in Section~\ref{sec:stability}. Statistical datasets and the map data downloaded from~\cite{natearth} are provided in the supplementary materials.

\subsection{Comparison to the State of the Art}\label{sec:soa}

\renewcommand{\figsize}{0.3}
\begin{figure*}[!hbt]
  \centering
  \includegraphics[height=5cm]{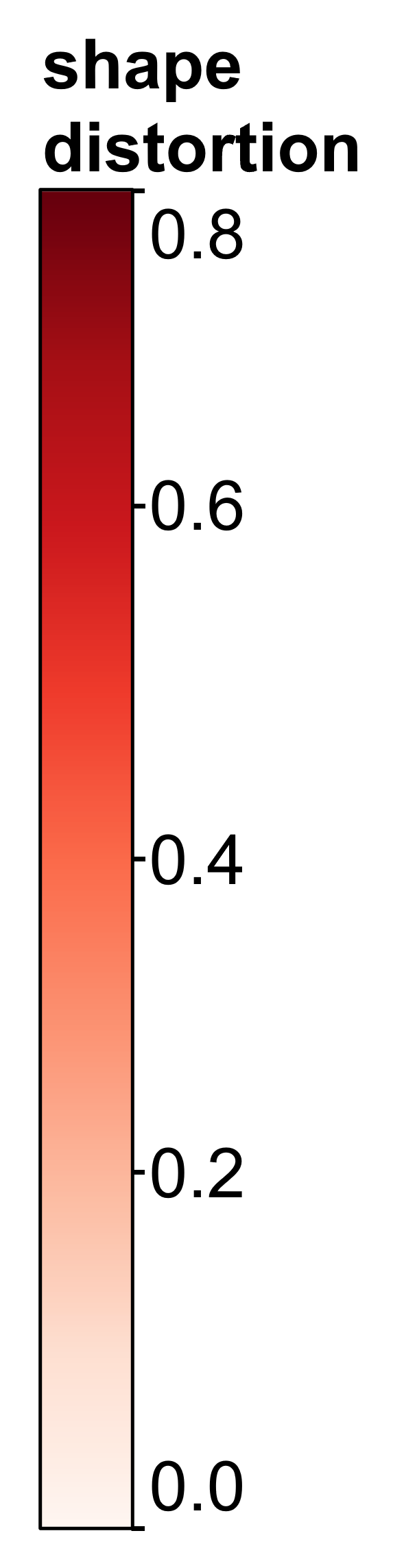}
  \hspace{-2mm}\subfloat[diffusion]{\includegraphics[width=\figsize\linewidth]{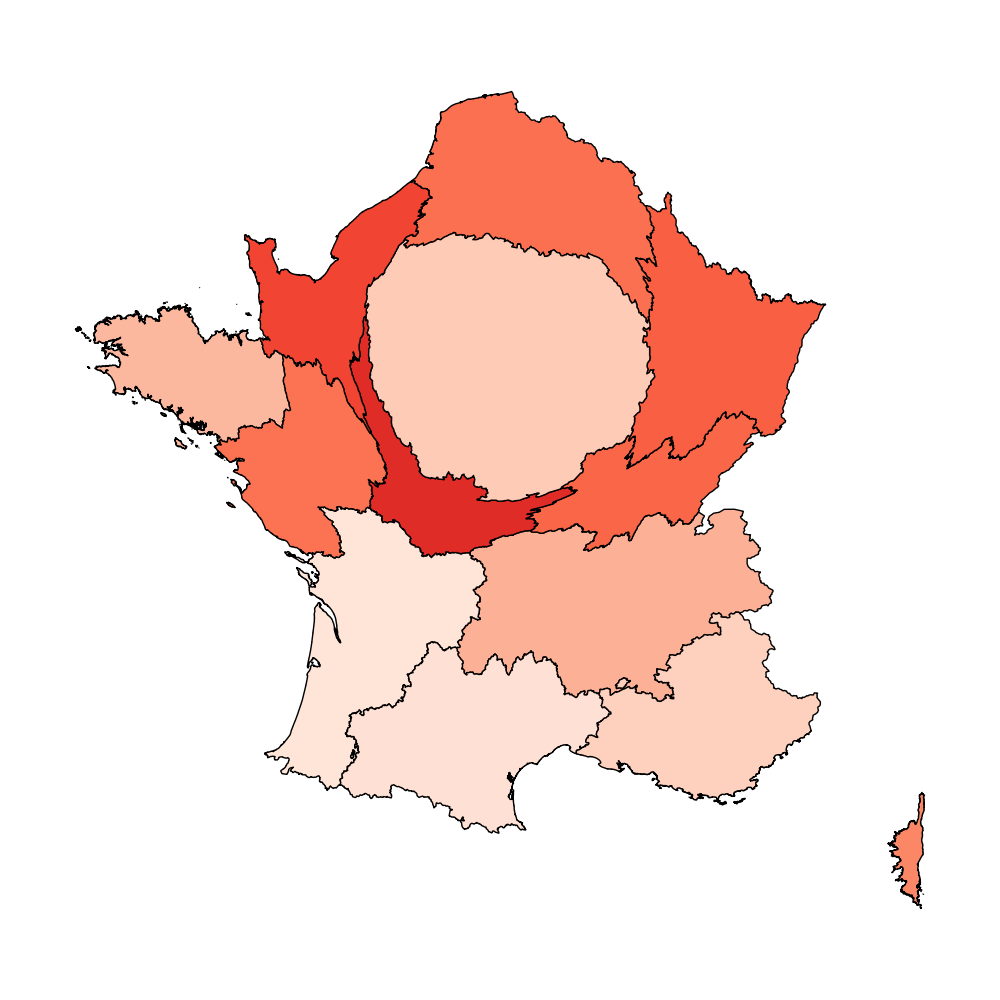}\label{fig::france_diffuse}}
  \hspace{-4mm}\subfloat[flow]{\includegraphics[width=\figsize\linewidth]{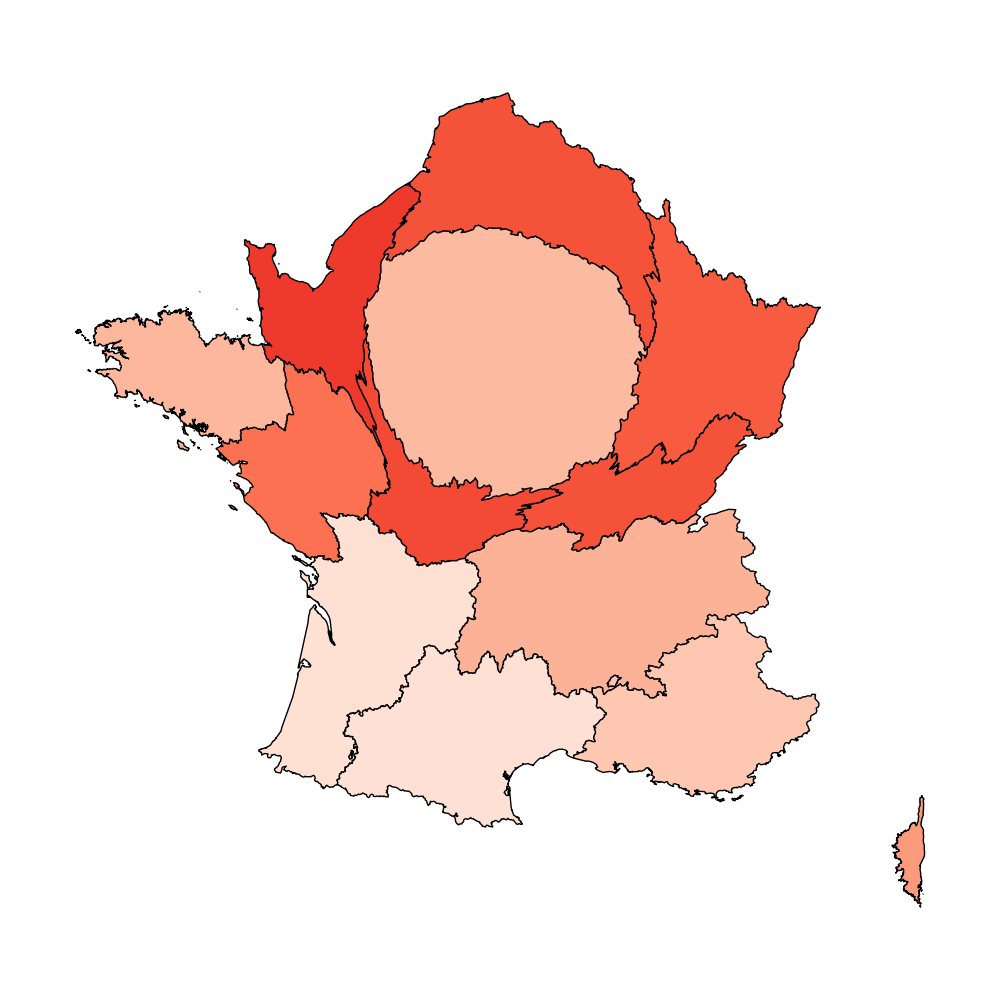} \label{fig::france_flow}}
  \hspace{-4mm}\subfloat[our]{\includegraphics[width=\figsize\linewidth]{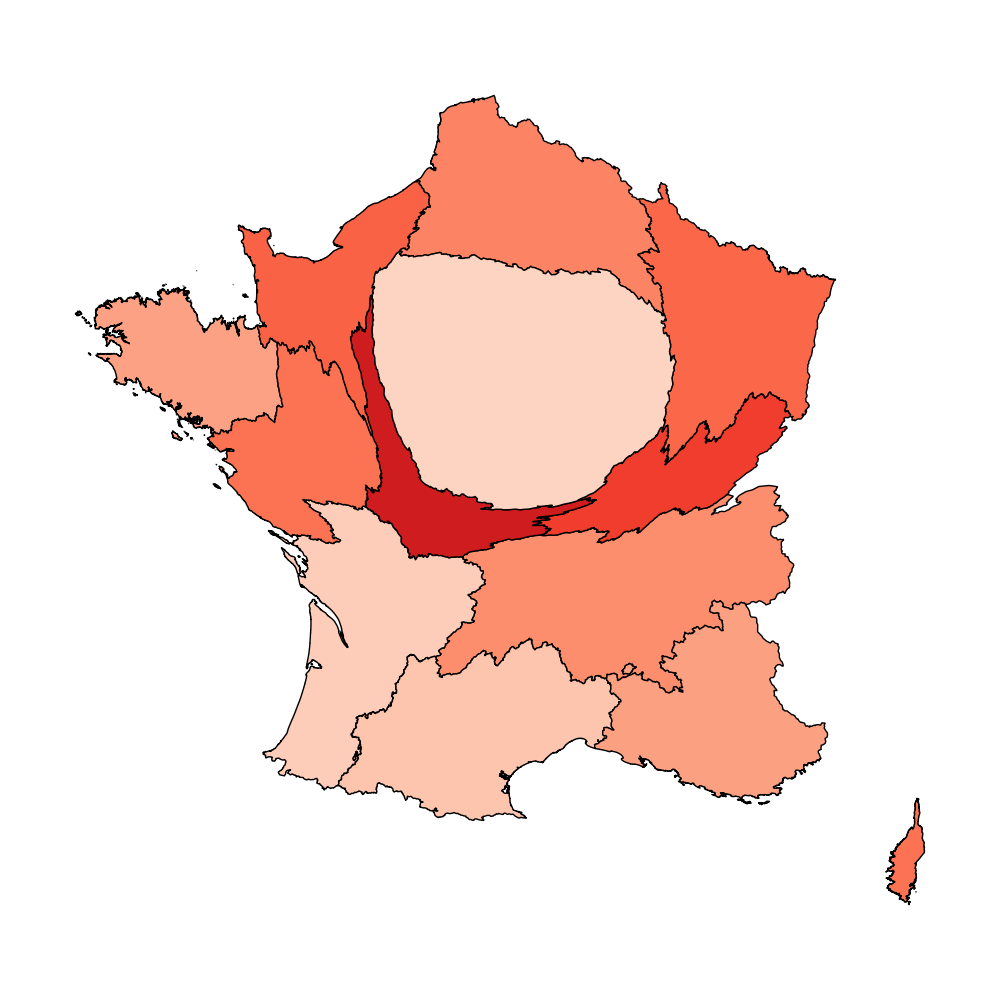}\label{fig::france_ours}}\\

  \subfloat[diffusion]{\includegraphics[width=\figsize\linewidth]{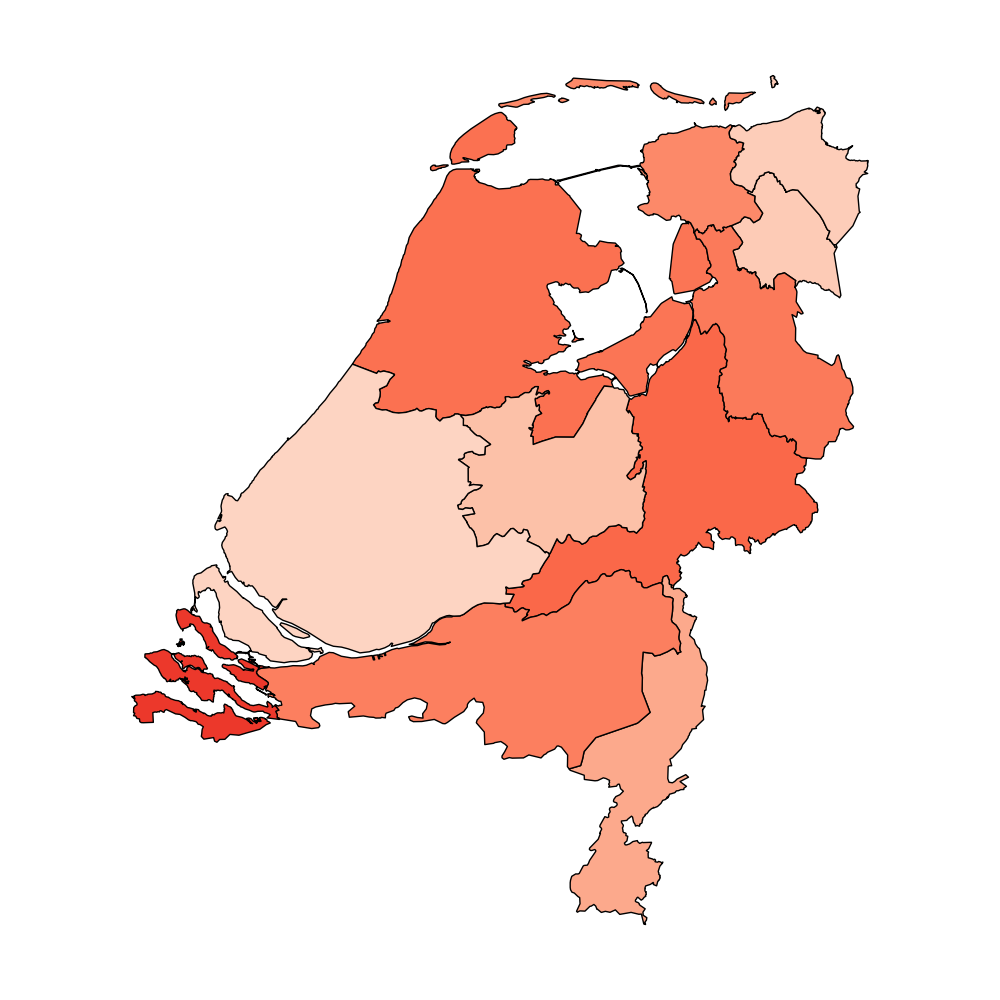}\label{fig::netherlands_diffuse}}
  \subfloat[flow]{\includegraphics[width=\figsize\linewidth]{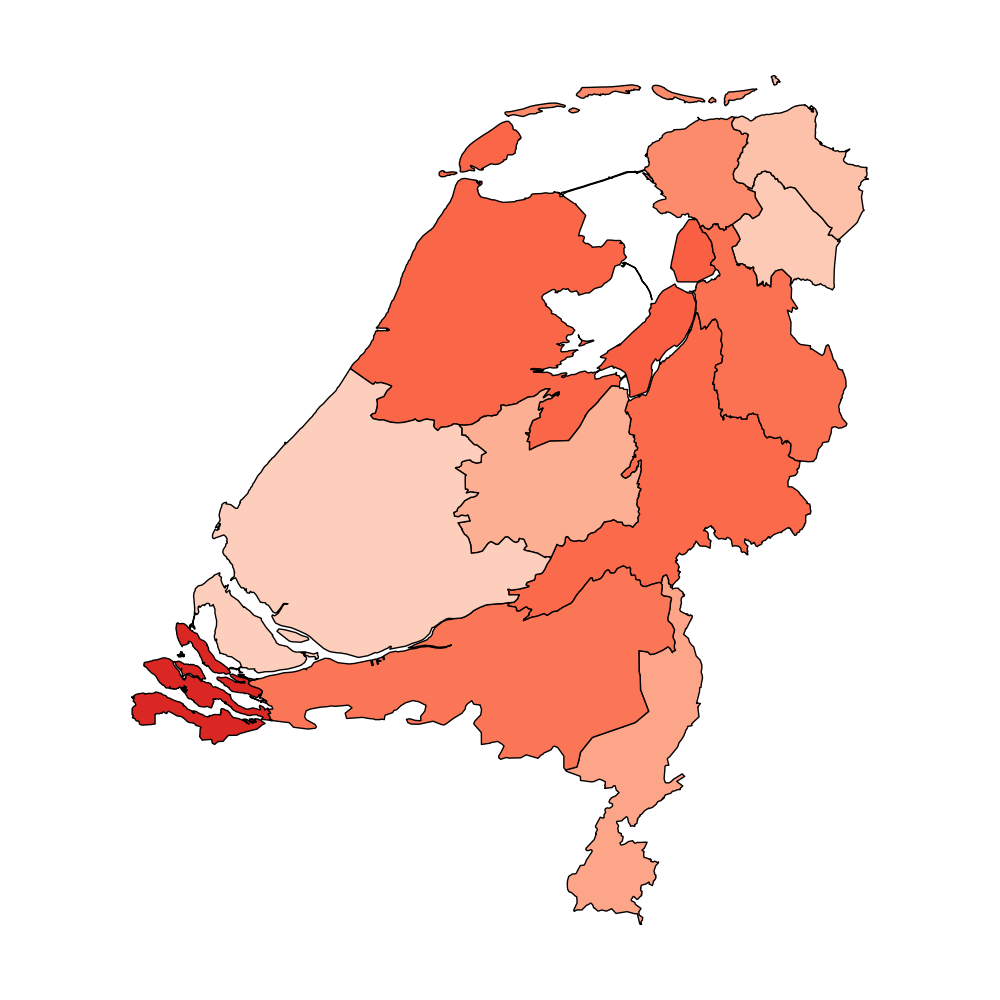} \label{fig::netherlands_flow}}
  \subfloat[our]{\includegraphics[width=\figsize\linewidth]{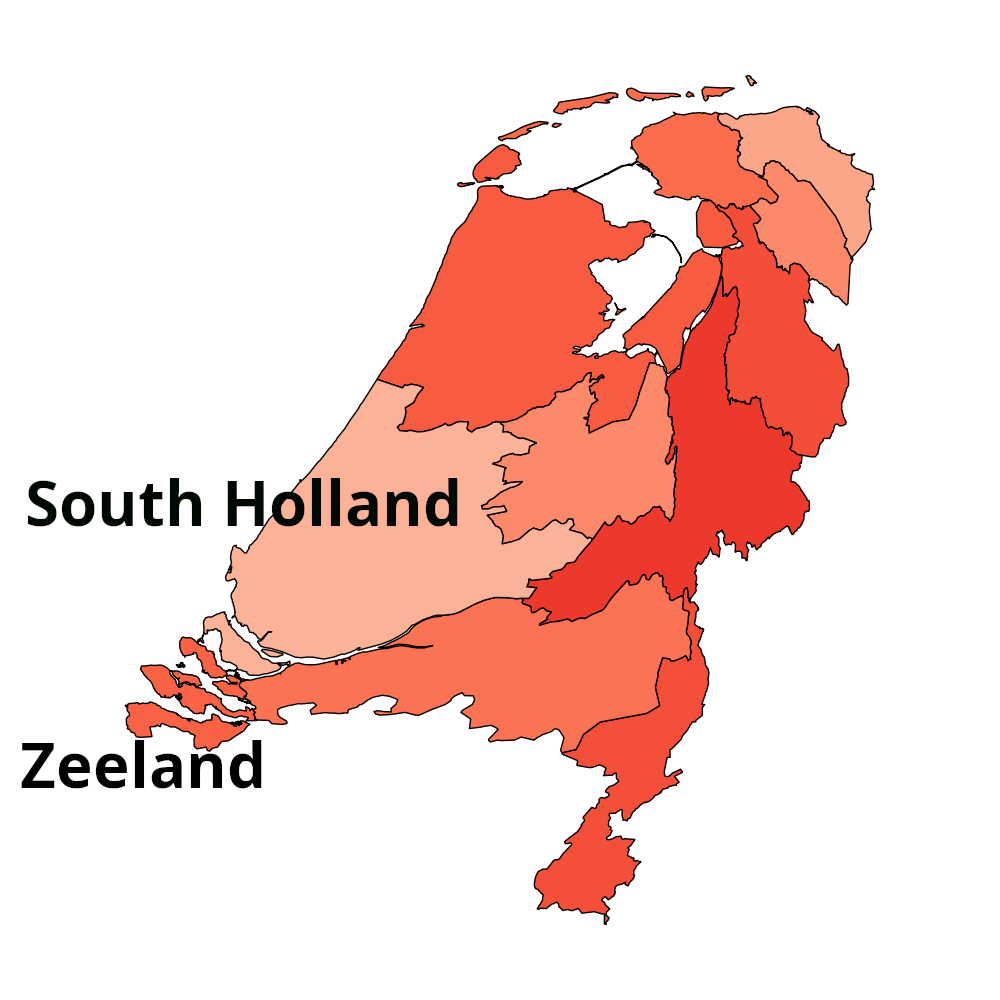}\label{fig::netherlands_ours}}\\

  \subfloat[diffusion]{\includegraphics[width=\figsize\linewidth]{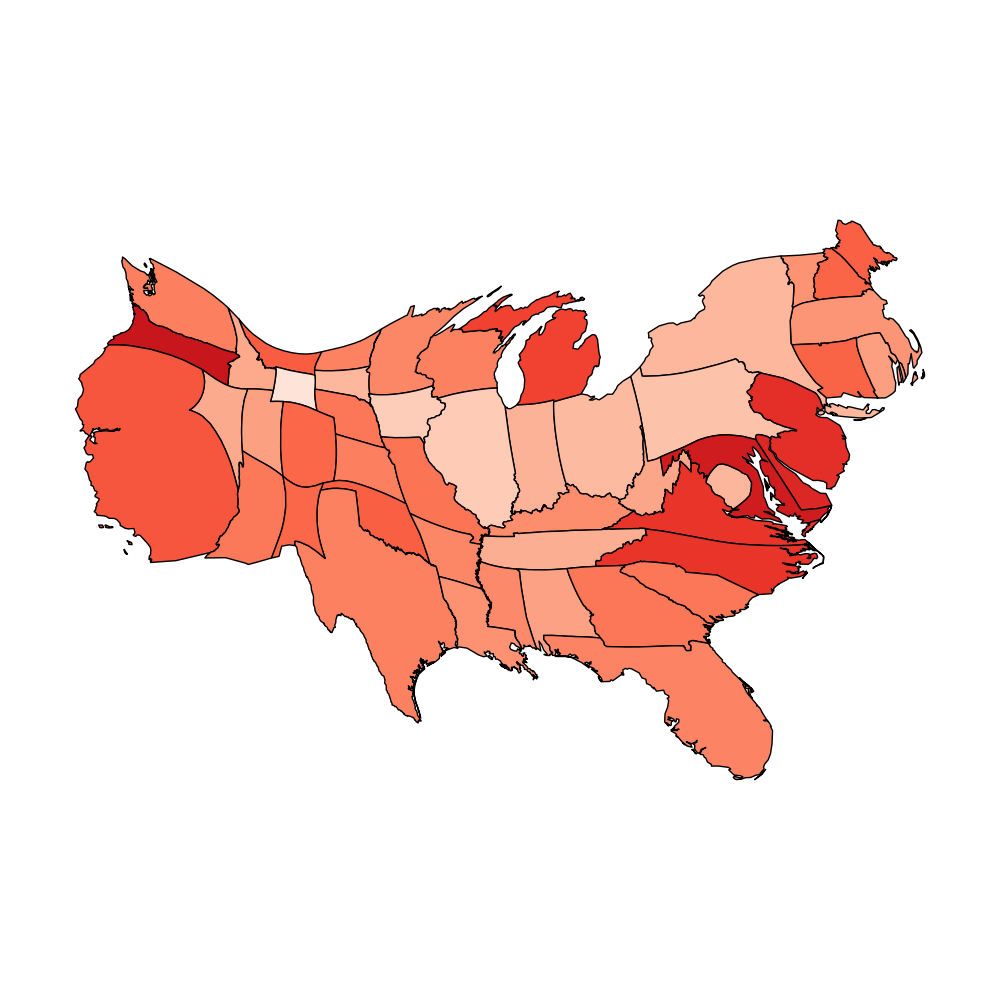}\label{fig::usa_diffuse}}
  \subfloat[flow]{\includegraphics[width=\figsize\linewidth]{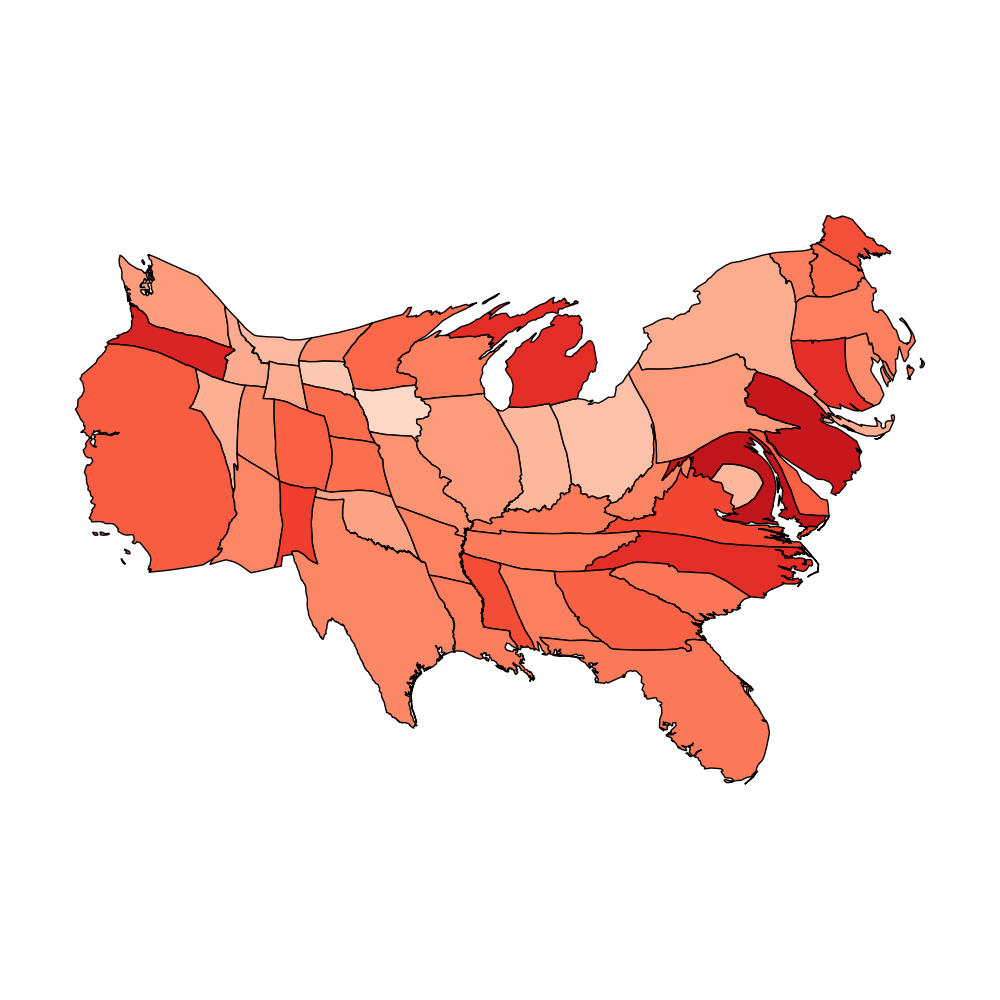} \label{fig::usa_flow}}
  \subfloat[our]{\includegraphics[width=\figsize\linewidth]{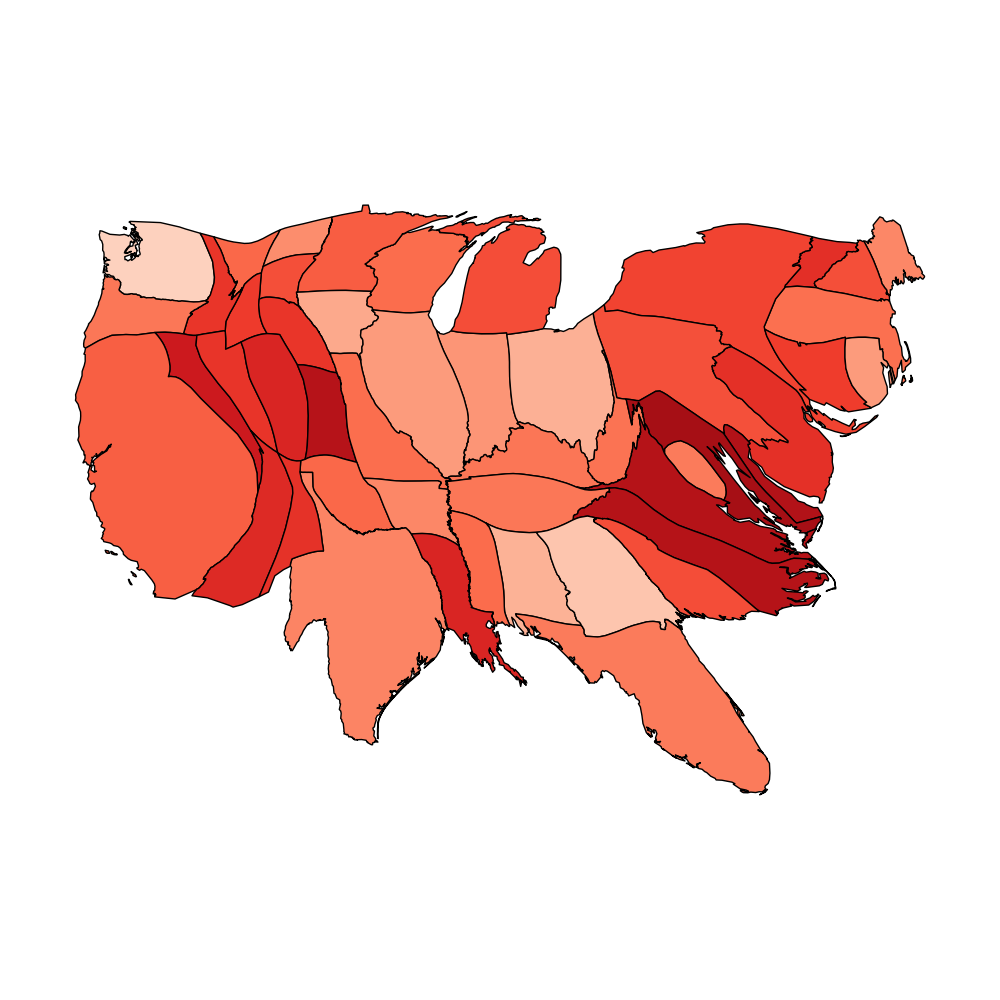}\label{fig::usa_ours}}\\

  \caption{Comparison of cartograms generated by diffusion~\cite{Gastner04}, flow-based~\cite{Gastner18}, and our method. Upper row: Population of France. Middle row: Population of the Netherlands. Lower row: Electoral votes in the USA. Color encodes shape distortion of regions.
  }
  \label{fig::qualitative}
\end{figure*}

For our tests, we used the datasets for the population of Germany, the Netherlands, and France, electoral votes in the USA, and GDP in Europe, which are available in~\cite{gocart}. The results of both physics-inspired cartogram methods are visually barely distinguishable. Differences from the maps generated with the proposed method are noticeable. For easier comparison, we color the regions according to their shape distortion. In all three examples, there exist regions with better shape preservation in the cartogram computed by our method. The resulting cartograms are easily readable for all methods, and all deformed regions can be related to their counterparts in the original map. The reader can find respective cartograms for Germany and Europe data in the supplementary materials.

We compare the results of the three approaches using six quality measures, see Figure~\ref{fig::comparison}. Note that cartographic errors below 0.01 are undetectable by visual inspection~\cite{Gastner18}. Our method reaches this level in all datasets except for the population of Europe, which can be related to the number of regions in the map. Our method demonstrates comparable values for the maximal distortion error and higher values for the average shape distortion and relative position errors.

\renewcommand{\figsize}{0.48}
\begin{figure}[!tbh]
  \centering
  \includegraphics[width=\figsize\linewidth]{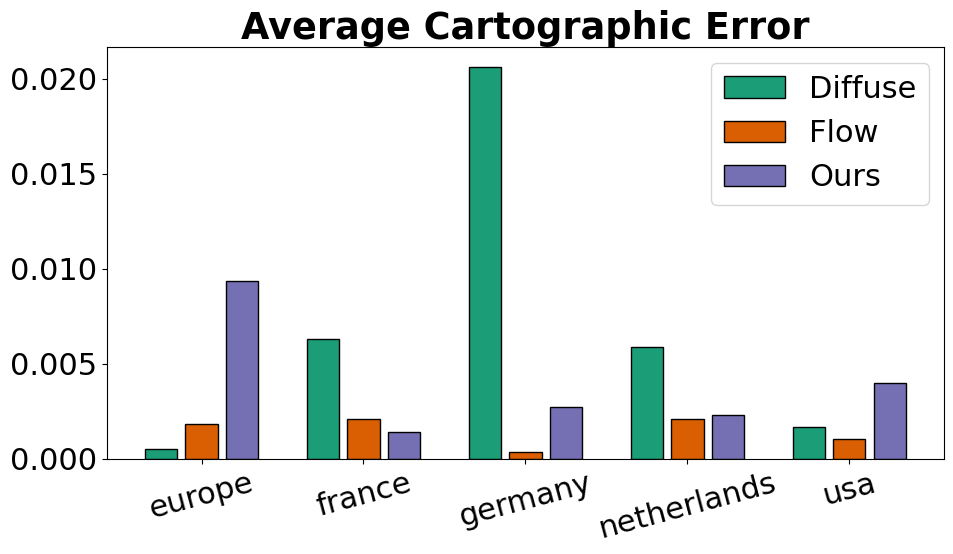}
  \includegraphics[width=\figsize\linewidth]{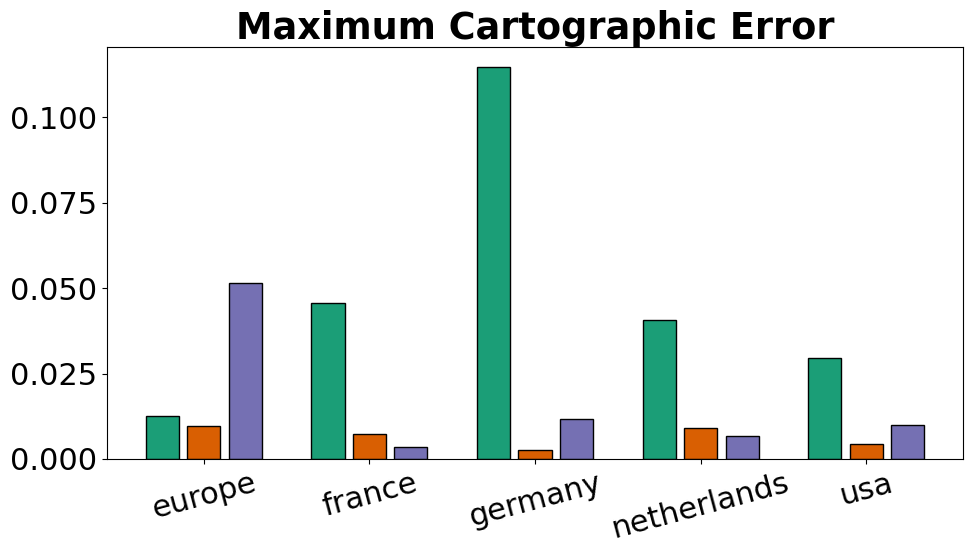}\\
  \includegraphics[width=\figsize\linewidth]{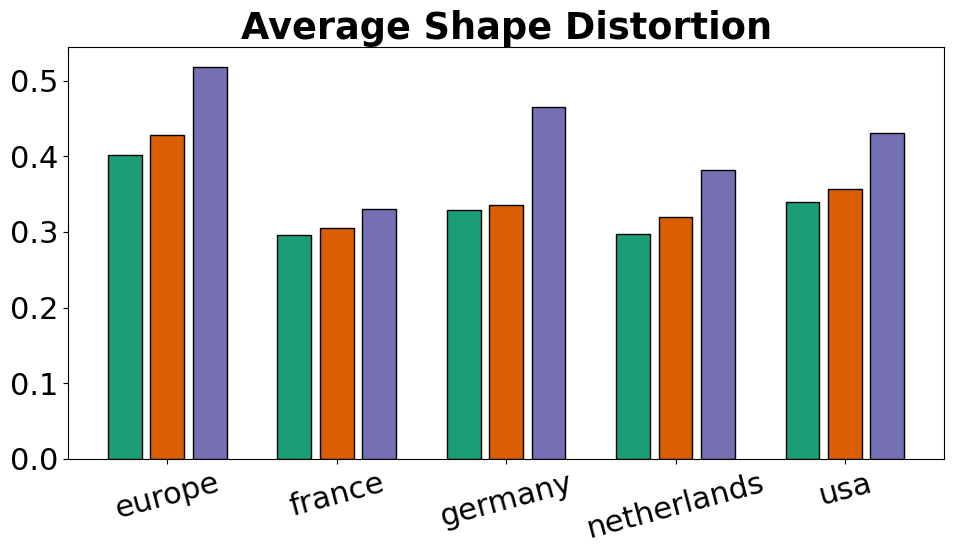}
  \includegraphics[width=\figsize\linewidth]{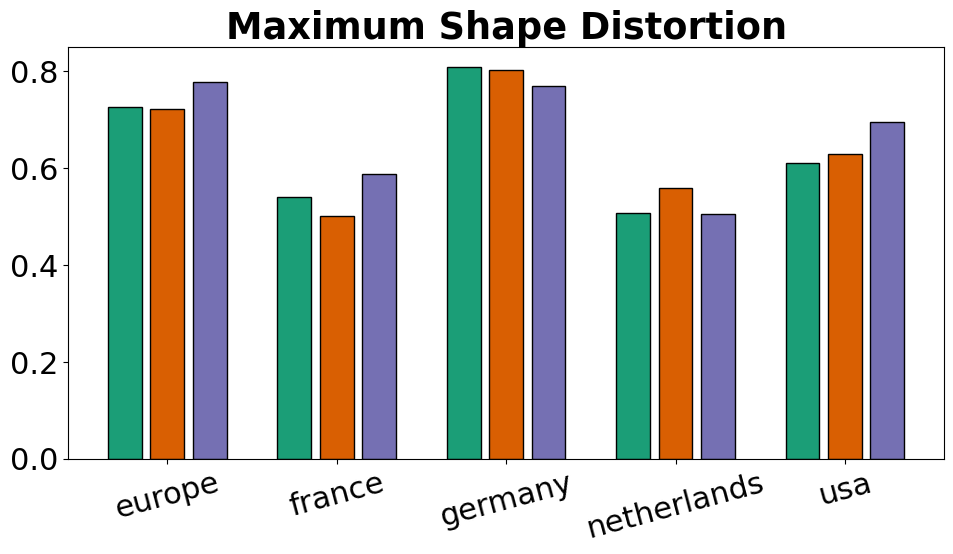}\\
  \includegraphics[width=\figsize\linewidth]{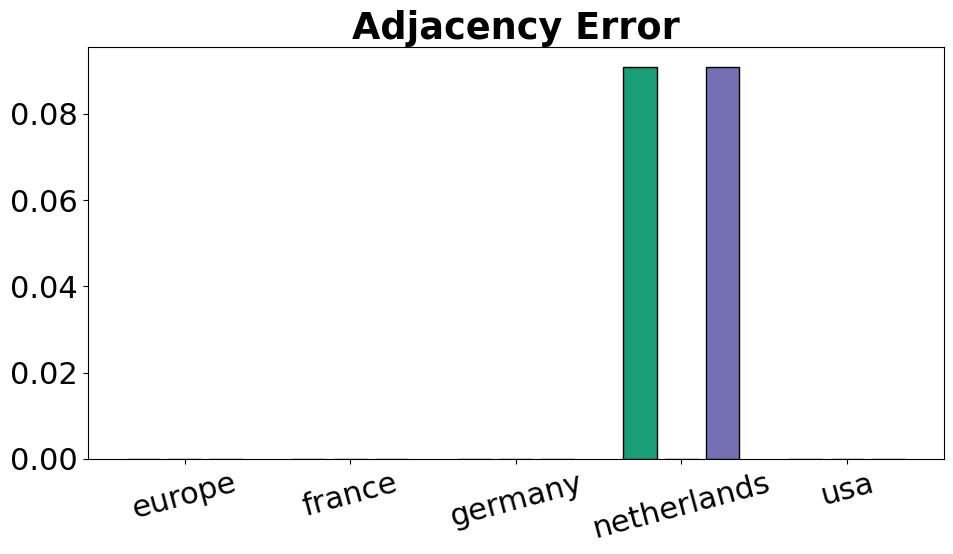}
  \includegraphics[width=\figsize\linewidth]{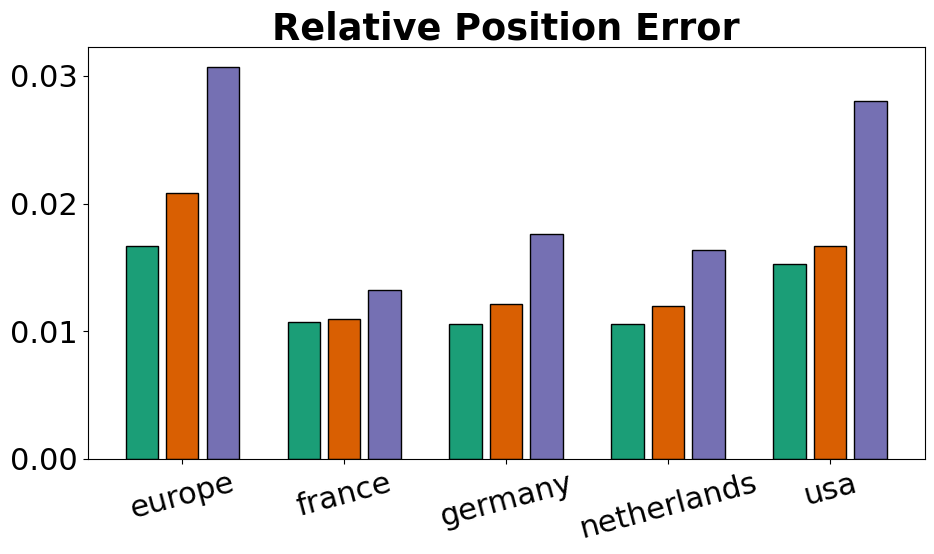}
  \caption{Quality measures for the cartograms presented in Figure~\ref{fig::qualitative}.
  }
  \label{fig::comparison}
\end{figure}

The highest cartographic errors for the Europe dataset were recorded for the regions of Iceland and the Republic of Cyprus, i.e., our method performs worse on isolated regions that are disconnected from the main set and are surrounded by the background. Note that the flow-based approach also demonstrates the highest cartographic error for Iceland. Topological errors are related to isolated regions as well. For example, the adjacency between Zeeland and South Holland in the cartogram of the Netherlands is violated. Figure~\ref{fig::iterations} presents the evolution of all quality measures over the iterations.

\renewcommand{\figsize}{0.48}
\begin{figure}[!tbh]
  \centering
  \includegraphics[width=\figsize\linewidth]{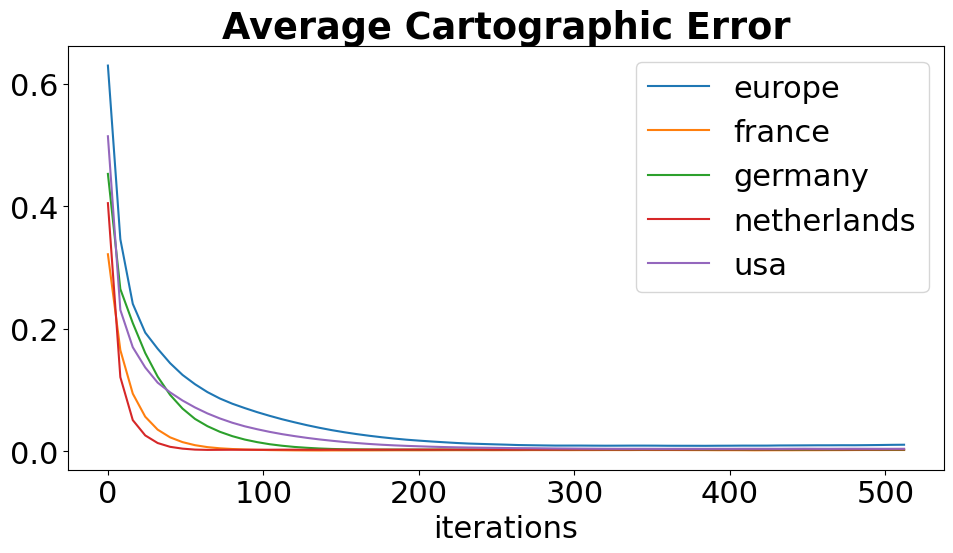}
  \includegraphics[width=\figsize\linewidth]{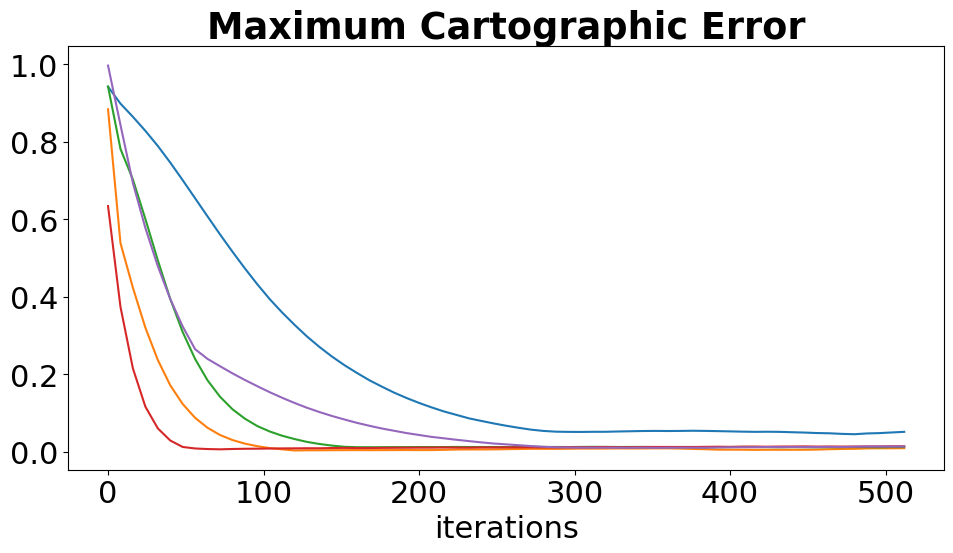}\\
  \includegraphics[width=\figsize\linewidth]{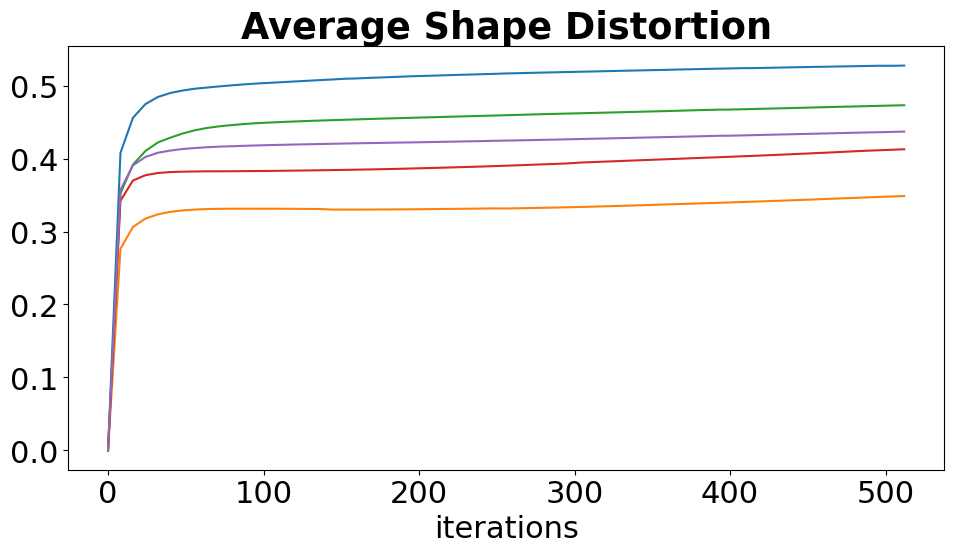}
  \includegraphics[width=\figsize\linewidth]{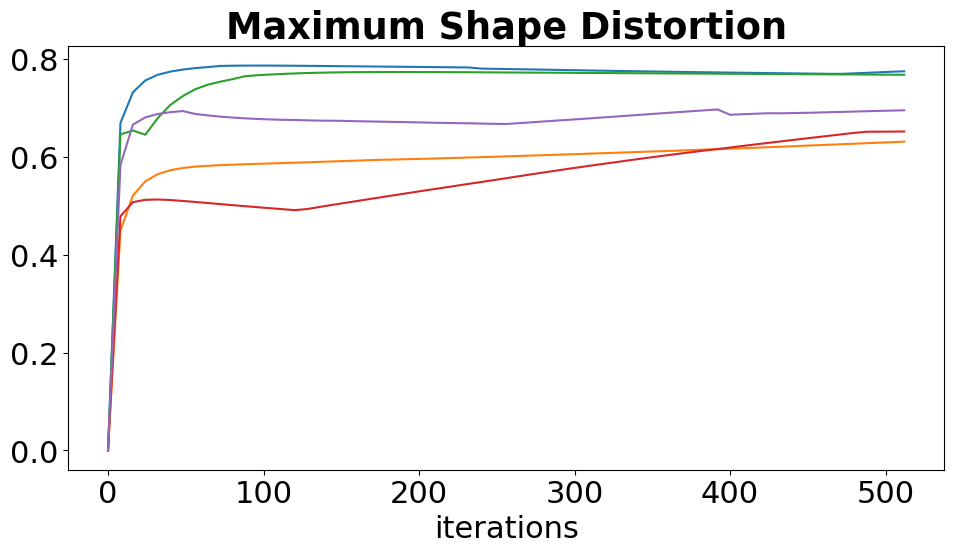}\\
  \includegraphics[width=\figsize\linewidth]{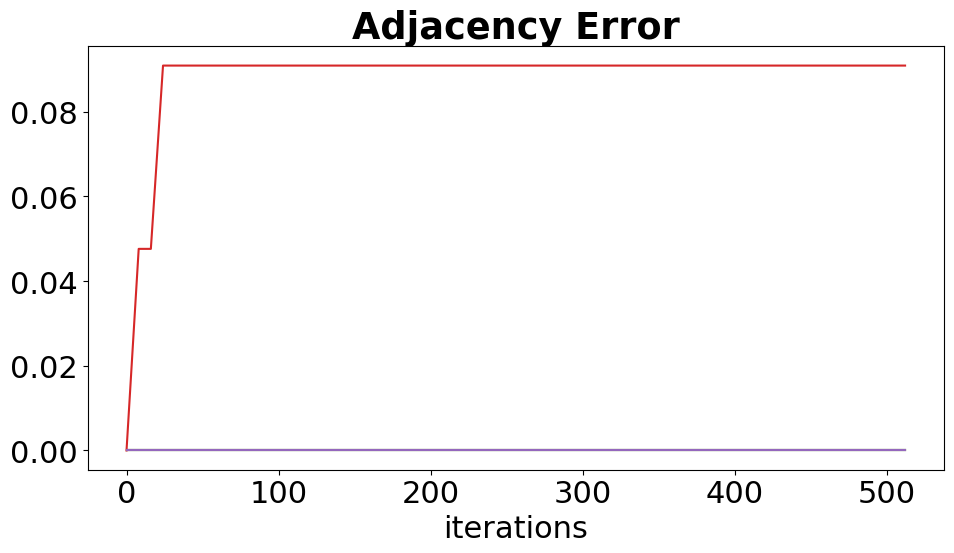}
  \includegraphics[width=\figsize\linewidth]{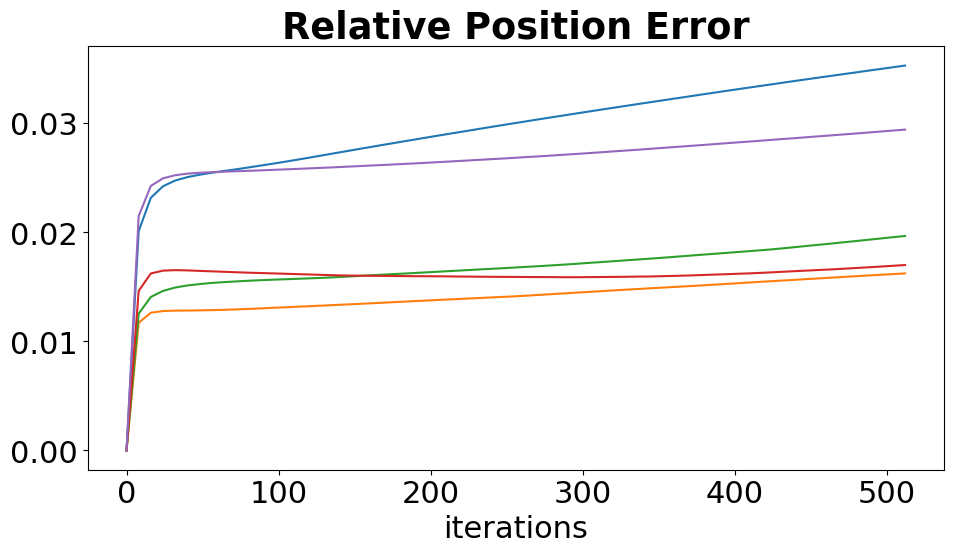}
  \caption{Evolution of quality measures for the cartograms presented in Figure~\ref{fig::qualitative}.
  }
  \label{fig::iterations}
\end{figure}

\subsection{Computational Performance}\label{sec:performance}
For typical input, the diffusion method finishes calculations within $10$ to $15$ seconds~\cite{gocart}. The novel flow-based algorithm~\cite{Gastner18} significantly outperforms the diffusion method. Therefore, we used the latter's performance measures as a reference for our tests. Numerical tests presented in this section were performed with the CPU i7-13620H and the GPU GeForce RTX 4070 Laptop.

Figure~\ref{fig::performance}~(left) shows computational times required by the proposed algorithm depending on the number of iterations. The slopes of the linear graphs depend on the complexity of geographical data. In particular, the execution time of the method grows as the number of regions and the total number of vertices of the polygonal region boundaries increase, see Table~\ref{tab:tab}. The algorithmical complexity of the proposed method is $\mathcal{O}(m+n)$, where $m=2^k\times2^k$ is the texture size and $n$ is the number of vertices in the polygonal region boundaries.

\renewcommand{\figsize}{0.495}
\begin{figure}[!hbt]
  \centering
  \includegraphics[width=\figsize\linewidth]{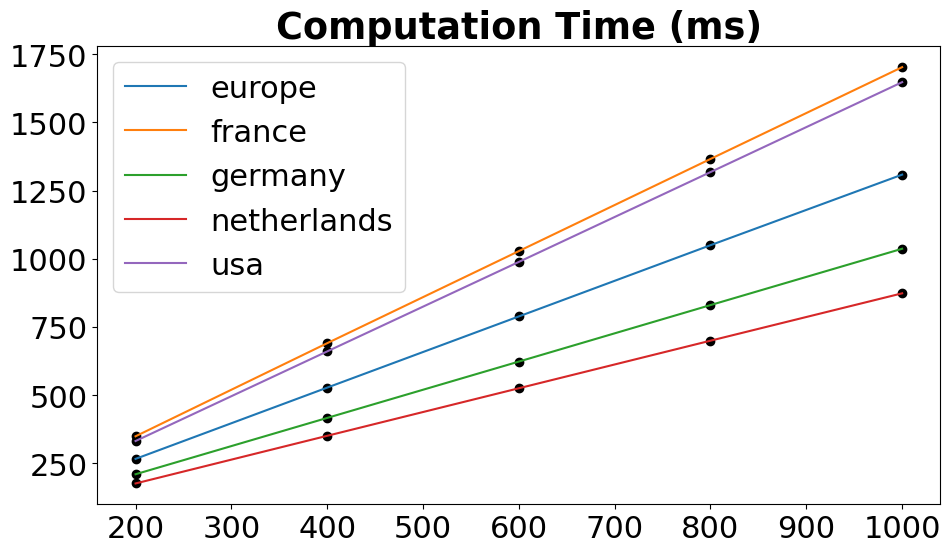}
  \includegraphics[width=\figsize\linewidth]{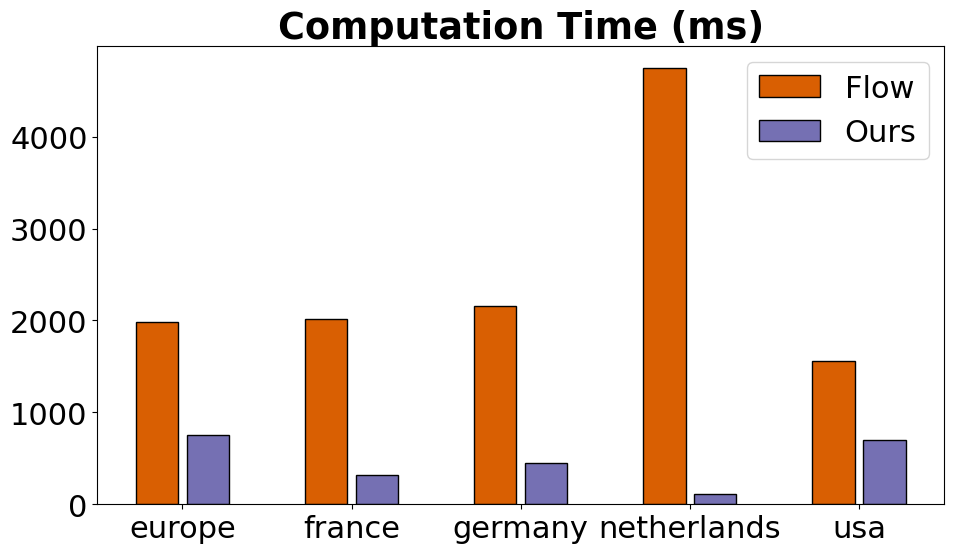}
  \caption{Performance measures of our proposed method for contiguous cartogram construction. Left:~computational time as a function of the number of iterations for five datasets. We observe a linear behavior. Right:~comparison of completion times of the flow-based algorithm and our proposed approach. Our algorithm is about one order of magnitude faster.}
  \label{fig::performance}
\end{figure}

\begin{table}[!hbt]
  \caption{Execution times (in ms) for one iteration of our proposed method when applied to different datasets. We also list the number of vertices and the number of regions of the different cartographic maps. }
  \label{tab:tab}
  \scriptsize%
  \centering%
  \begin{tabu}{%
      r%
        *{5}{c}%
    }
    \toprule
    dataset           & Europe & France & Germany & Netherlands & USA   \\
    \midrule
    num. of vertices  & 6283   & 53885  & 18320   & 4961        & 11904 \\
    num. of regions   & 35     & 13     & 16      & 12          & 49    \\
    one iteration, ms & 1.32   & 1.72   & 1.04    & 0.88        & 1.65
  \end{tabu}%
\end{table}

\subsection{Initial Scaling of the Map}\label{sec:scaling}

The flow-based method by Gastner et al.~\cite{Gastner18} uses the fast Fourier transform to compute distortions. The vanishing flow condition is enforced at the boundaries of the visualization domain. To reduce the influence of boundary shapes on the resulting cartogram, the authors recommended using a texture with a side length equal to $1.5$ times the linear extent of the original map. In our numerical experiments, we utilized a considerably thinner boundary layer. This approach reduces computational effort on less relevant parts of the map while enabling higher resolution in areas containing statistical data. However, the thinner boundary layer may adversely affect the relative position and shape distortion of the regions.

Figure~\ref{fig::scaling} illustrates the results of our numerical tests evaluating quality measures as a function of the scaling of the original map. A scaling factor of $0 < s < 1$ indicates that the undeformed map is scaled to fit within the interval $[0.5-\frac{s}2, 0.5+\frac{s}2]$ in its largest dimension (horizontal or vertical). We observe that larger values of $s$ correspond to smaller cartographic errors. This can be attributed to improved resolution of map regions at higher scaling factors. However, other quality measures generally increase with larger $s$, meaning that distortions of the original map become more pronounced as less empty space is available in the visual domain.

\renewcommand{\figsize}{0.48}
\begin{figure}[!tbh]
  \centering
  \includegraphics[width=\figsize\linewidth]{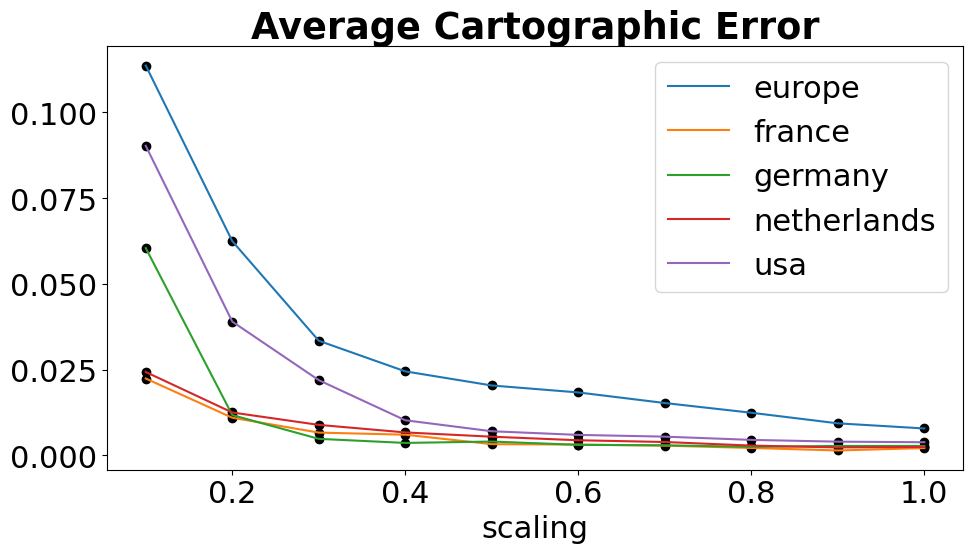}
  \includegraphics[width=\figsize\linewidth]{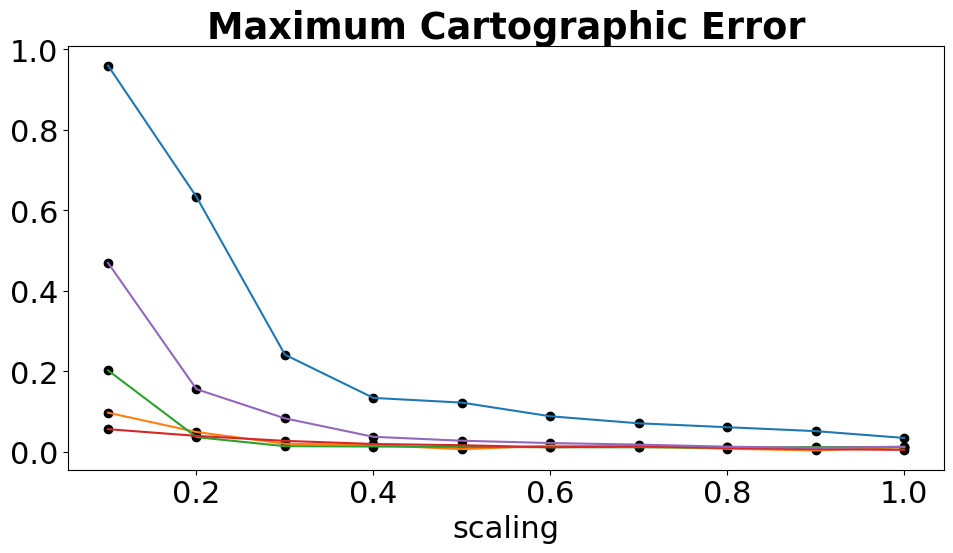}\\
  \includegraphics[width=\figsize\linewidth]{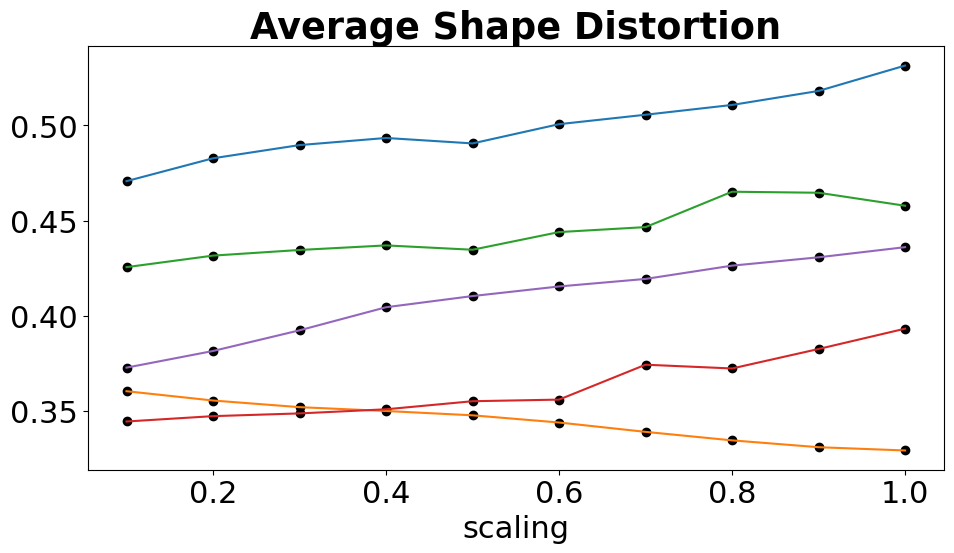}
  \includegraphics[width=\figsize\linewidth]{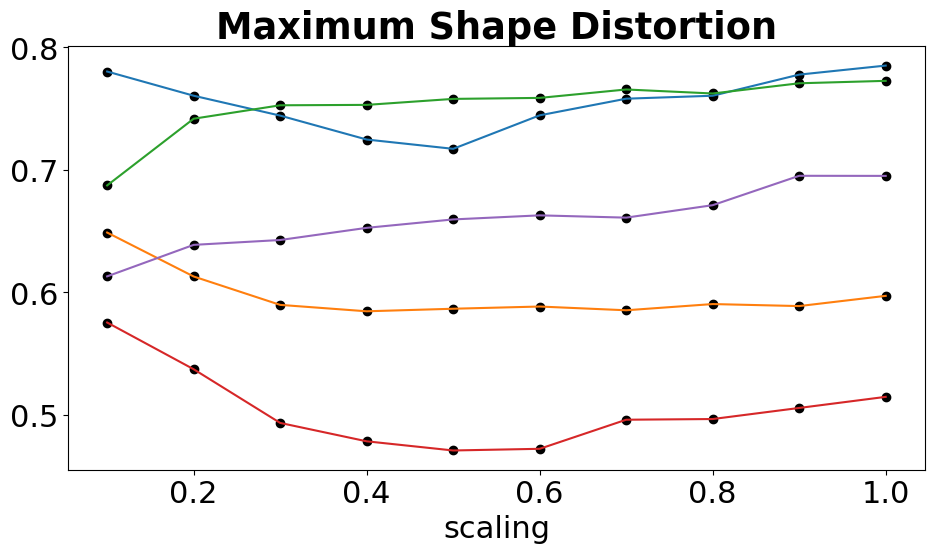}\\
  \includegraphics[width=\figsize\linewidth]{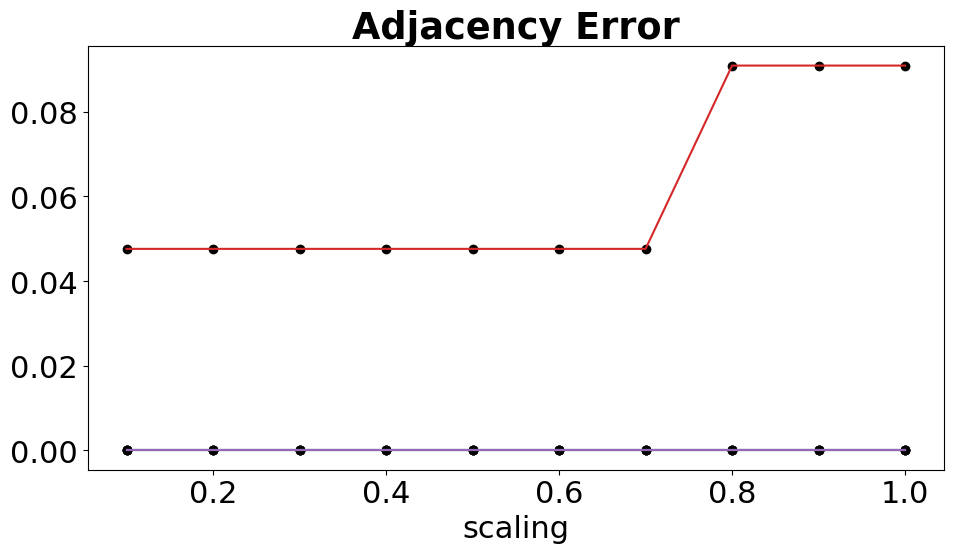}
  \includegraphics[width=\figsize\linewidth]{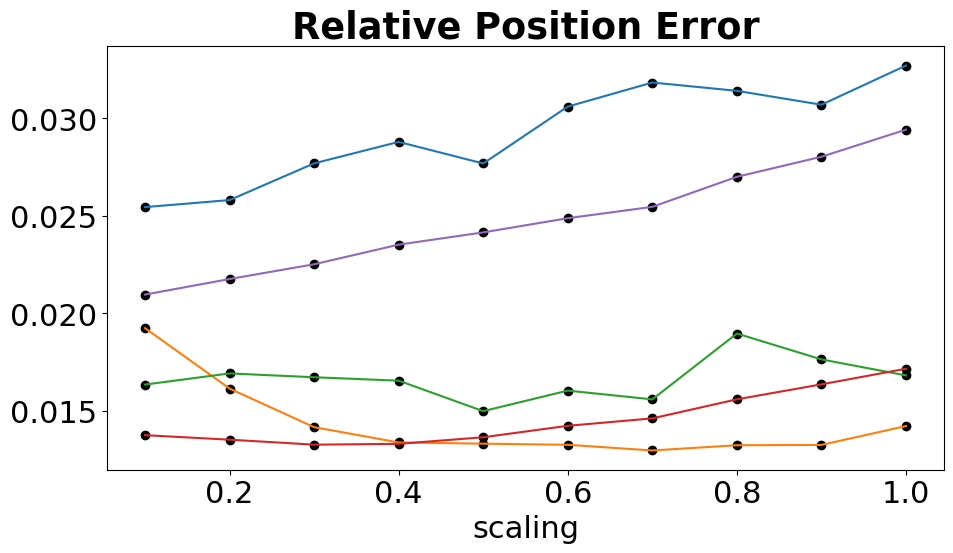}
  \caption{Effect of scaling on quality measures for the cartograms presented in Figure~\ref{fig::qualitative}.}
  \label{fig::scaling}
\end{figure}

\subsection{Background Density Value}\label{sec:backvalue}

The background is the part of the visualization domain that does not belong to any region, thus, no statistical data are provided. The lack of statistical data raises the question of how to choose the density function value for the background. Vanishing density would lead to a dramatic shrinking of the background in the cartogram. That effect would also lead to significant shape and topology distortions, which reduce the quality of the cartogram. A reasonable choice is to set the density in the background region equal to the average density ${d_0}$ computed over all other regions. This choice provides a certain resistance of the background area to shape and topology distortions.

The quality measures introduced in Section~\ref{sec:quality} allow a numerical evaluation of the choice of the BDV. Figure~\ref{fig::back_density} shows how these measures vary when the assigned density deviates up to $\pm 5\%$ from the average value ${d_0}$. In all five datasets, the behavior of all five quality measures shows that ${d_0}$ is the most reasonable choice for the BDV.

Changing the background density is also possible in the flow-based method by Gastner et al.~\cite{Gastner18}. The authors propose the idea of using spatially varying BDV to enhance shape preservation. However, this approach was not implemented to maintain the simplicity of the algorithm. Our proposed method, instead, allows the user to interactively control the spatially uniform background density with no effect on the computational complexity of the method.

\renewcommand{\figsize}{0.48}
\begin{figure}[!hbt]
  \centering
  \includegraphics[width=\figsize\linewidth]{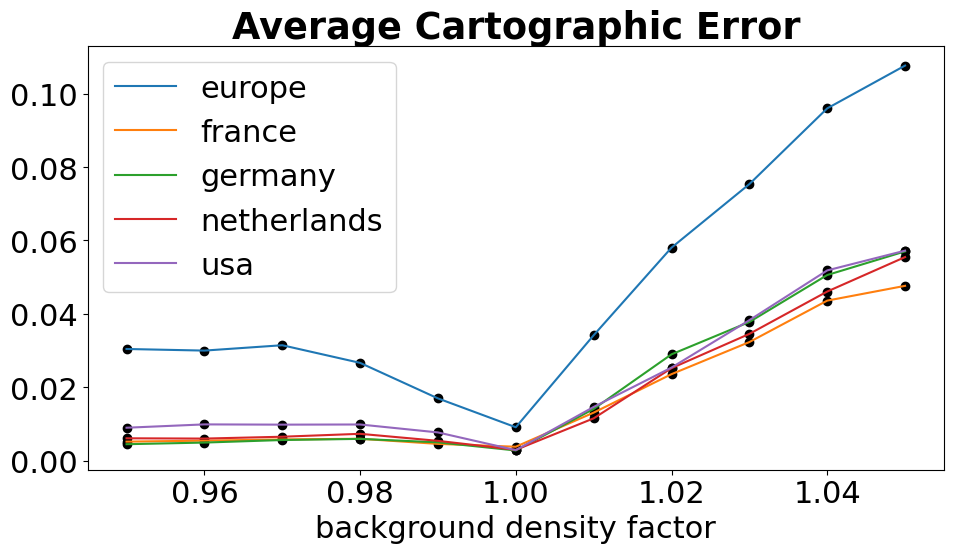} \label{fig::bdv_avr_error}
  \includegraphics[width=\figsize\linewidth]{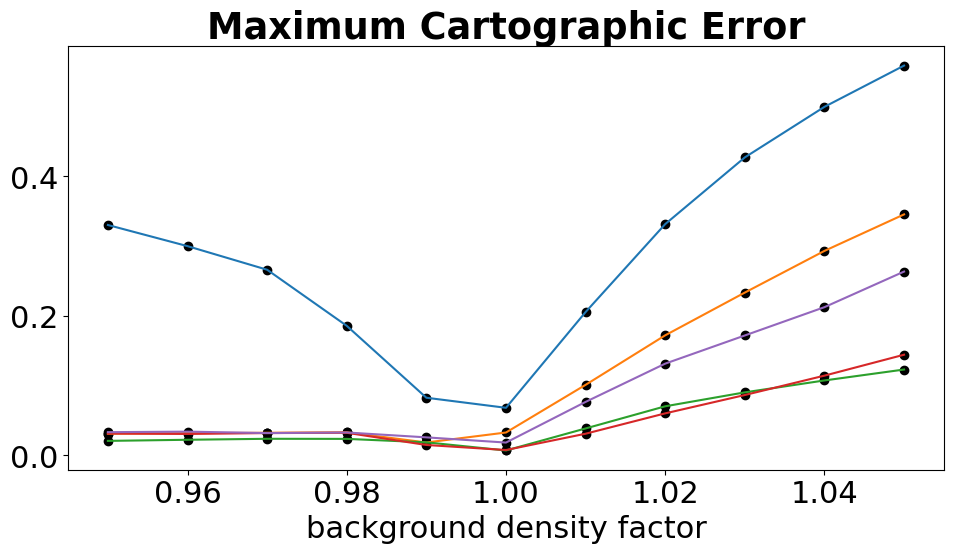}\label{fig::bdv_max_error}\\
  \includegraphics[width=\figsize\linewidth]{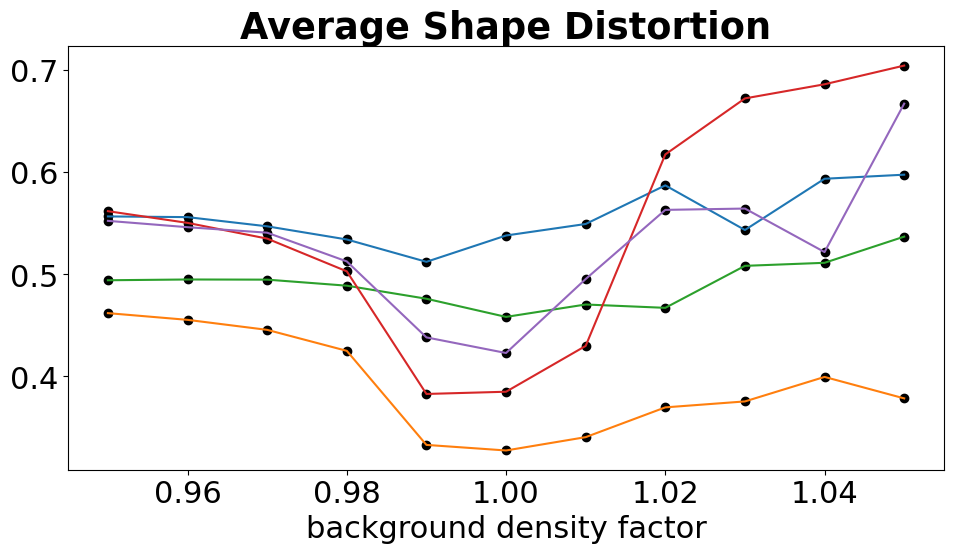} \label{fig::bdv_avr_shape}
  \includegraphics[width=\figsize\linewidth]{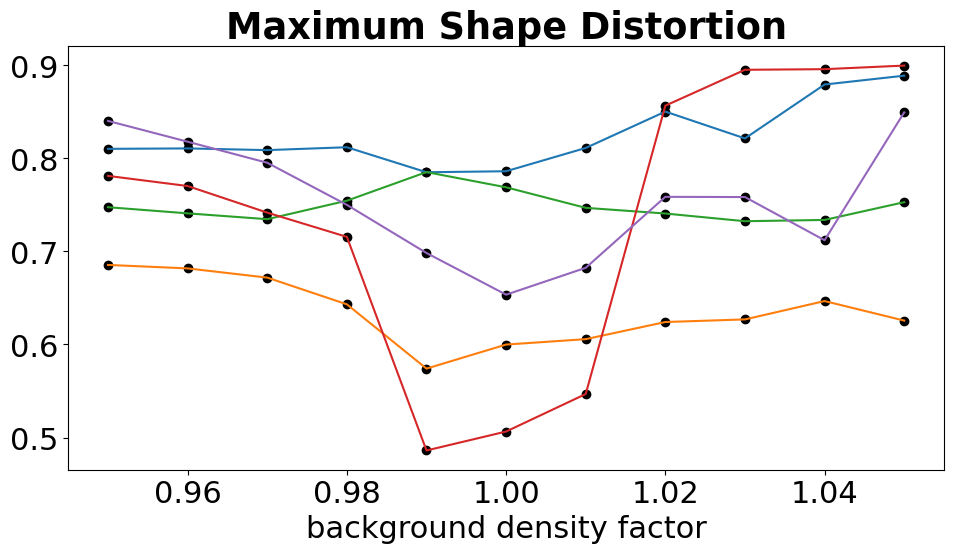} \label{fig::bdv_max_shape}\\
  \includegraphics[width=\figsize\linewidth]{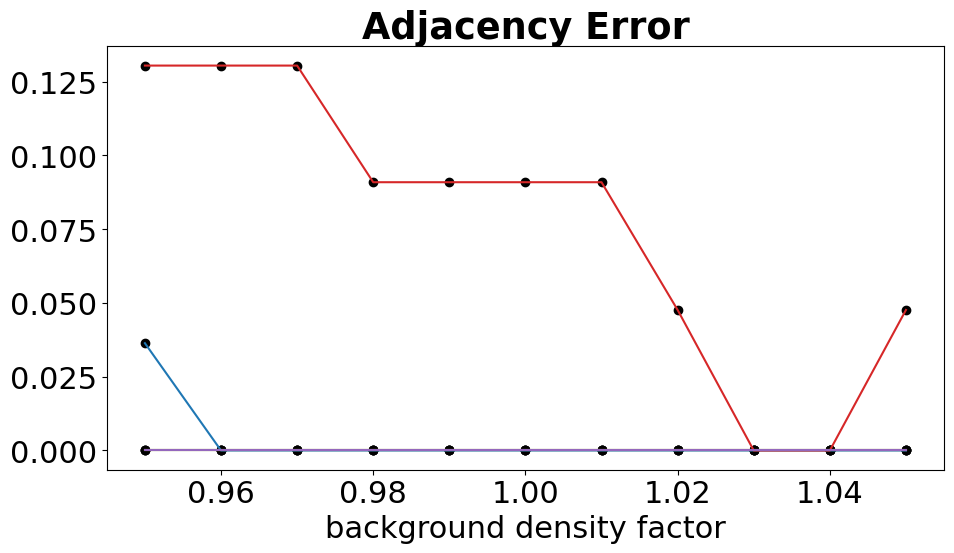} \label{fig::bdv_adjecency}
  \includegraphics[width=\figsize\linewidth]{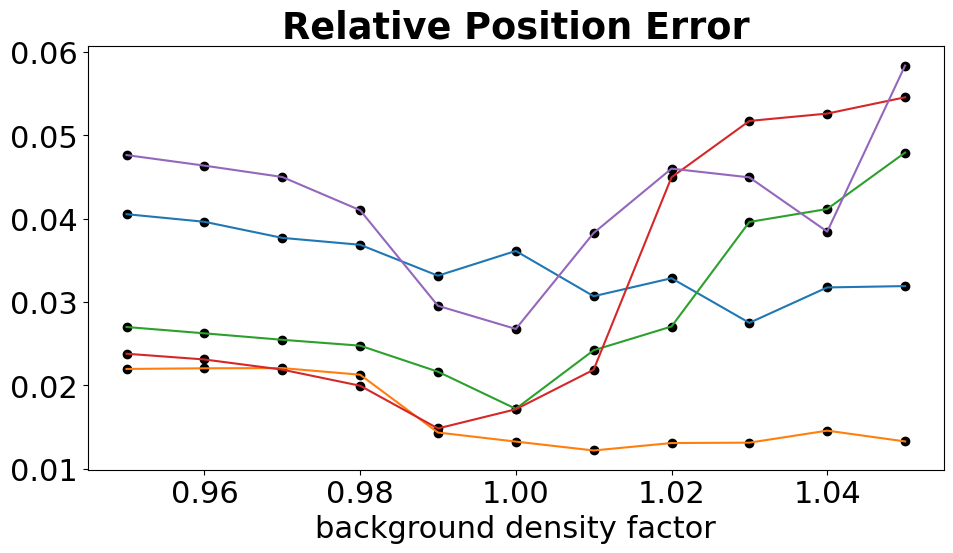} \label{fig::bdv_position}\\
  \caption{Dependence of different types of errors on the values of background density. Value $1.00$ on the horizontal axes corresponds to the BDV equal to ${d_0}$, i.e., the average density value over the map regions. Deviations of up to $\pm 5\%$ from ${d_0}$ were considered in the tests. All tests indicate that ${d_0}$ is the optimal choice for the BDV.
  }
  \label{fig::back_density}
\end{figure}

\subsection{Direct vs Cumulative Approaches}\label{sec:stability}

When generating cartograms for time-varying data, computational efficiency and stability become important characteristics of the algorithm. Here, stability is understood as the similarity of the cartogram layouts computed for similar statistical data~\cite{Nickel22} or, equivalently, the coherence of cartograms constructed for smoothly varying temporal data.

We demonstrate the stability of our method using data representing the dynamics of the reported number of weekly COVID-19 cases by state in the United States~\cite{weekly_covid_usa}. As the weekly reported data vary, these changes are reflected and can be tracked by the user in the dynamic cartogram layout. In our first test, we perform $200$ iterations of the DET to compute cartograms for each time step starting from the original map, i.e., we applied the direct approach. The computation of one cartogram takes approximately $0.5$ seconds. The optimal BDV was chosen for each time step individually, as discussed in Section~\ref{sec:backvalue}. A few snapshots are shown in Figure~\ref{fig:dynamic_cartogram}. The figures demonstrate the similarity of the cartograms' layouts computed for moderately changed data. The persistence of the state locations allows for studying the dynamics of their individual statistics over time. The resulting animation is shown in the accompanying video, in which we generated $10$ frames per time step by linearly interpolating the underlying cartographic data.

\renewcommand{\figsize}{0.485}
\begin{figure}[!bth]
  \centering
  \includegraphics[width=\figsize\linewidth]{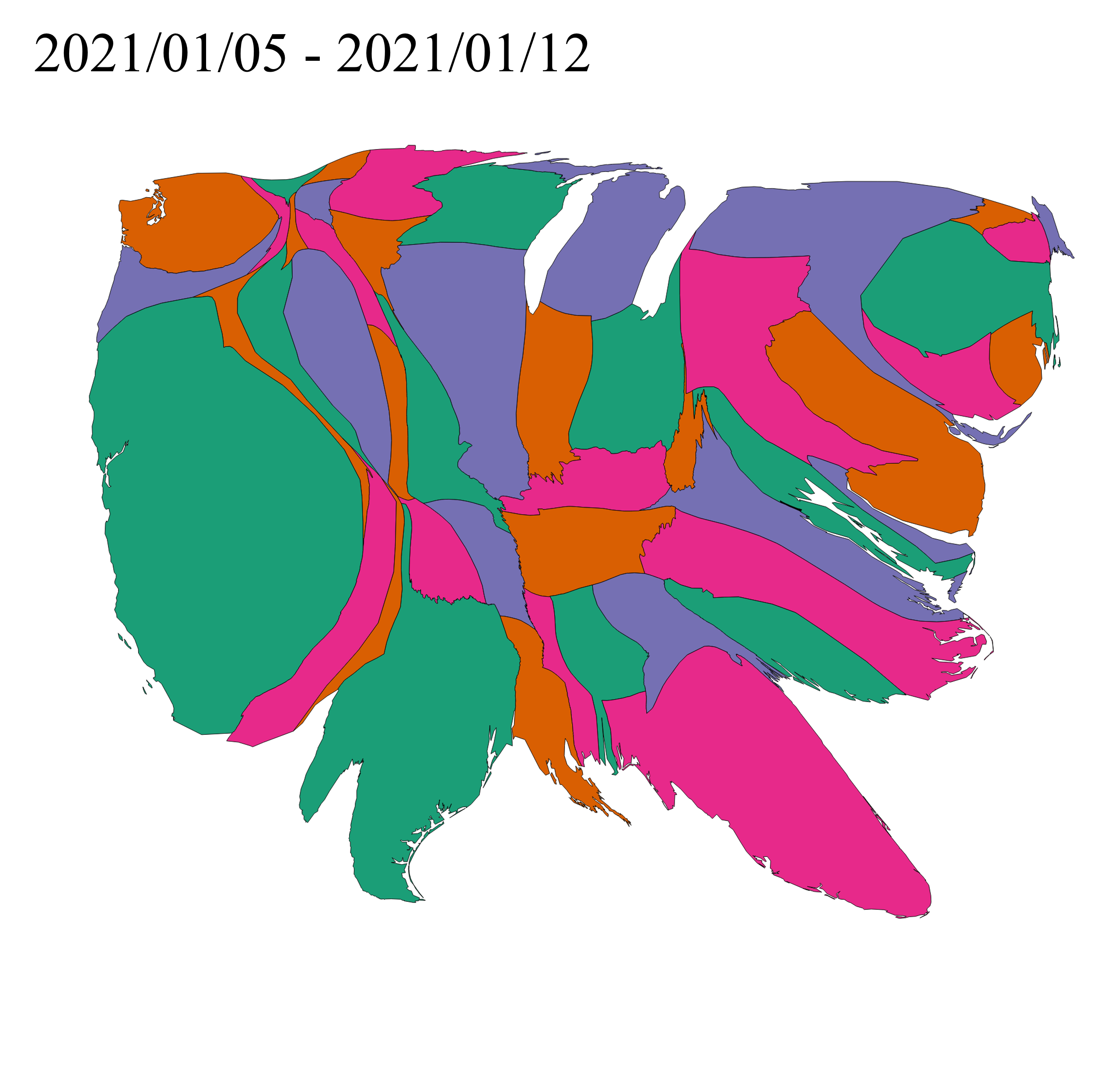}
  \includegraphics[width=\figsize\linewidth]{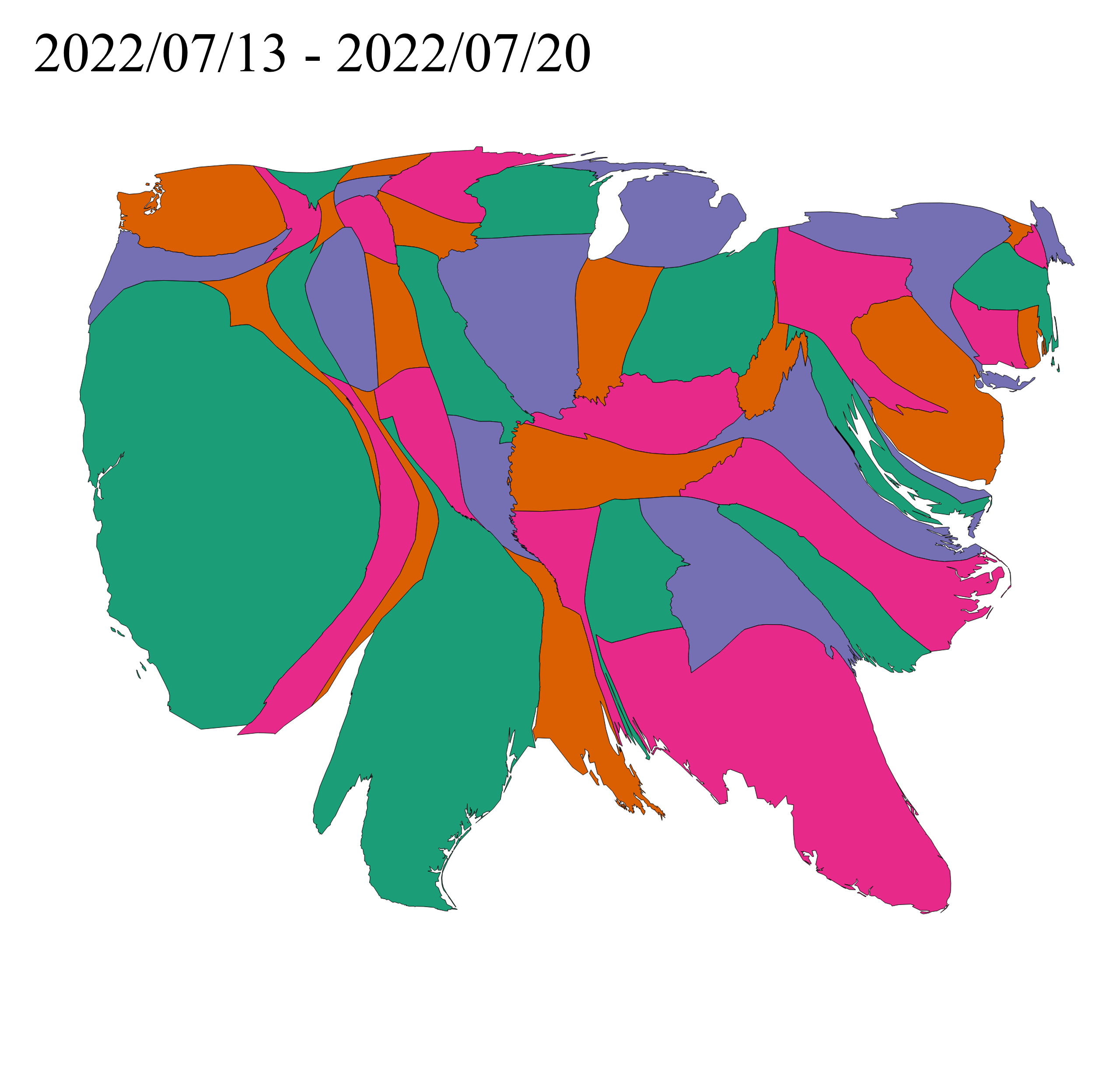}\\ \vspace{-2em}
  \caption{Weekly United States new COVID-19 cases (dataset from \cite{weekly_covid_usa}). The persistence of the cartograms' layout allows for visual comparison of the cartograms' regions. The cartograms were computed using the direct approach.}\label{fig:dynamic_cartogram}
\end{figure}

In our next experiment, we progressively deformed the map, i.e., we reused the cartogram from the previous time step to compute a new cartogram. We use identical criteria for stopping the iterations and the same optimal BDVs as in the first test. The resulting cartograms representing the number of weekly \textit{new} COVID-19 cases in the United States (dataset from \cite{weekly_covid_usa}) are shown in Figure~\ref{fig:direct_vs_cumulative}. We notice that, at time step $40$, the cumulative cartogram marginally differs from the one computed by the direct approach. However, $80$ steps later, the cumulative cartogram becomes overly distorted.
We provide a plot of the dynamics of the integral statistics $M_i(t)$, which shows that the cumulative approach fails when data vary significantly.

\renewcommand{\figsize}{0.485}
\begin{figure}[!bth]
  \centering
  \includegraphics[width=0.95\linewidth]{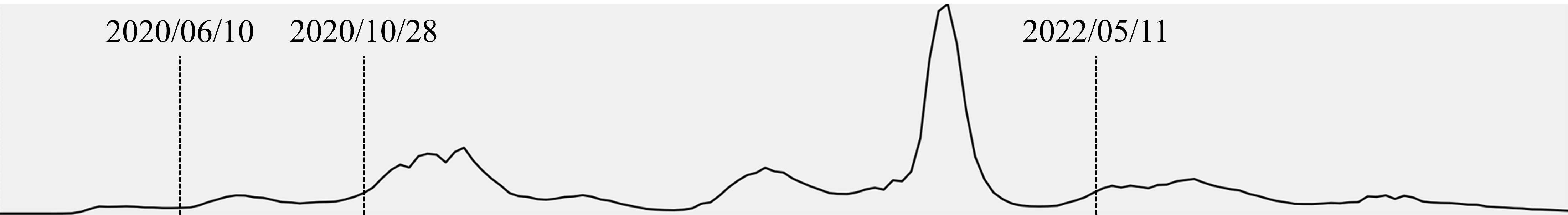}\\
  \includegraphics[width=\figsize\linewidth]{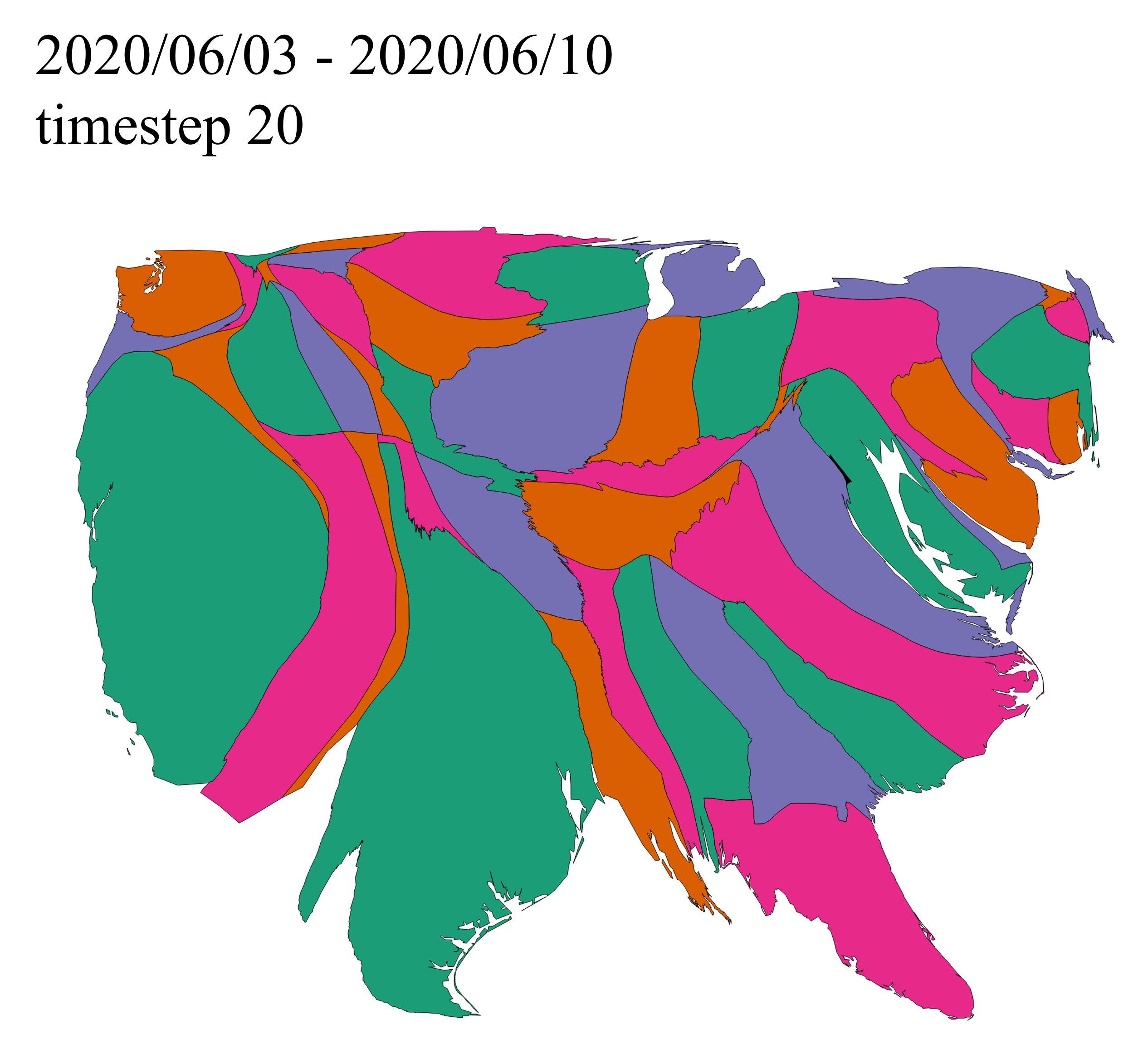}
  \includegraphics[width=\figsize\linewidth]{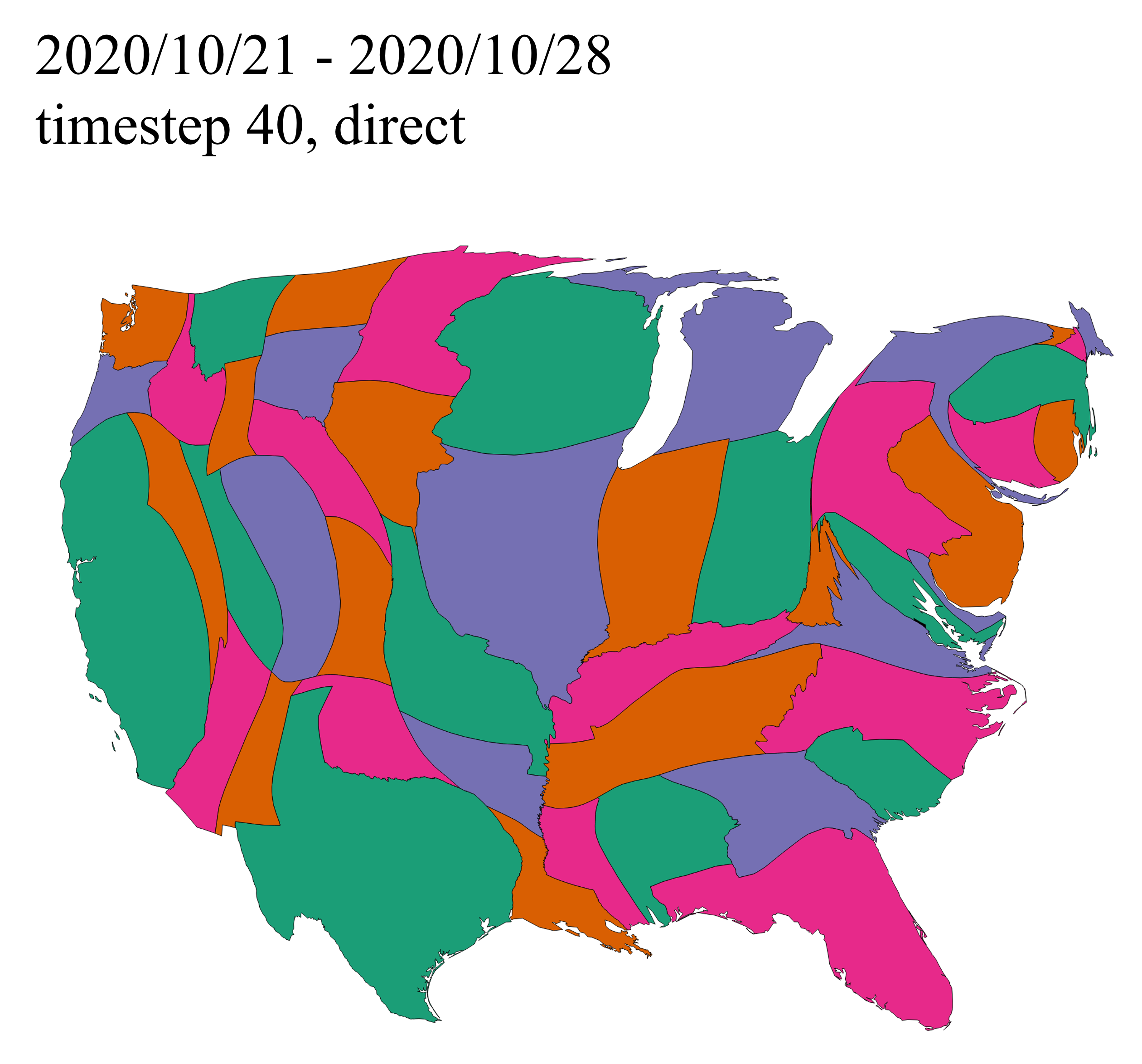}\\
  \includegraphics[width=\figsize\linewidth]{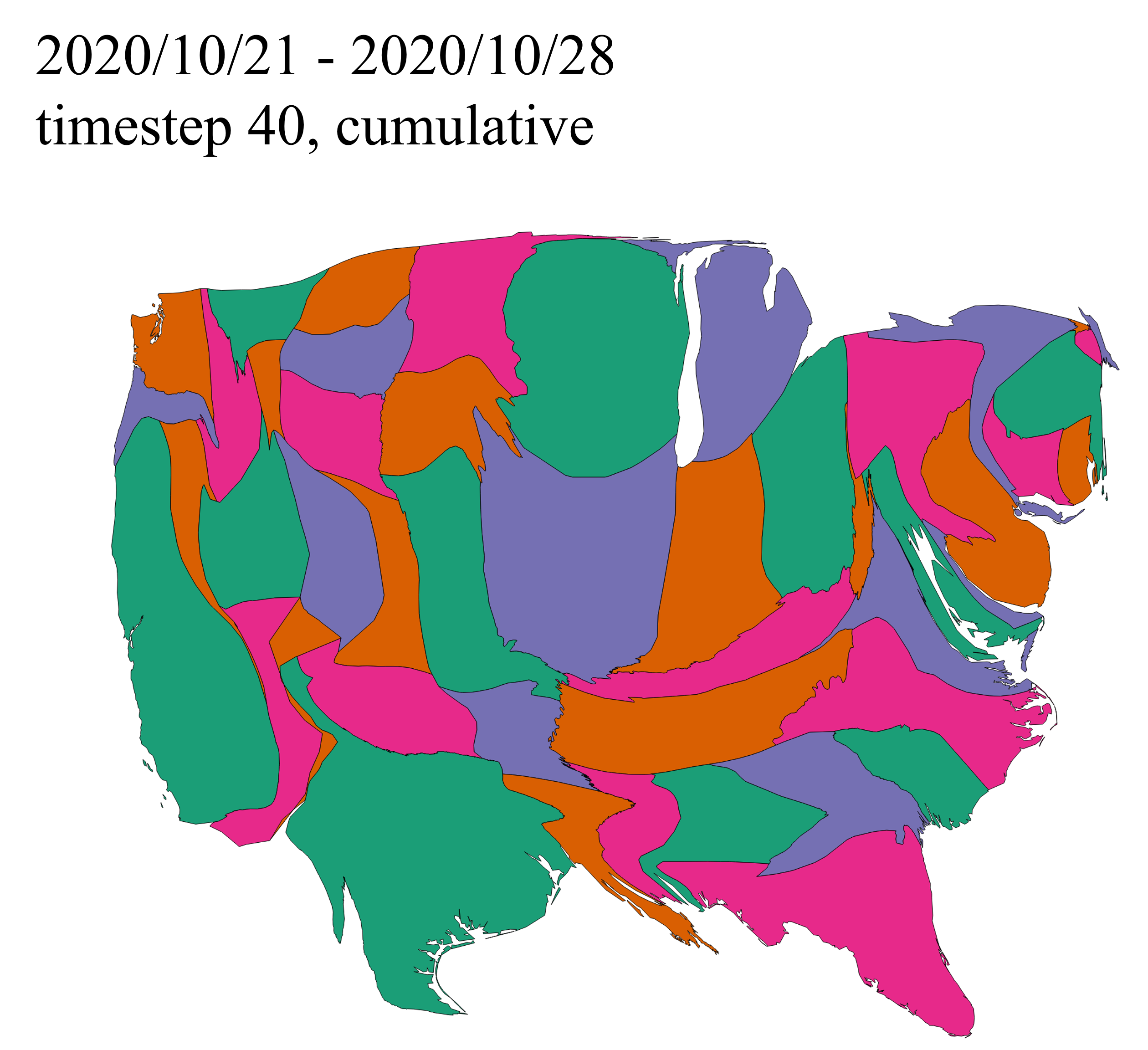}
  \includegraphics[width=\figsize\linewidth]{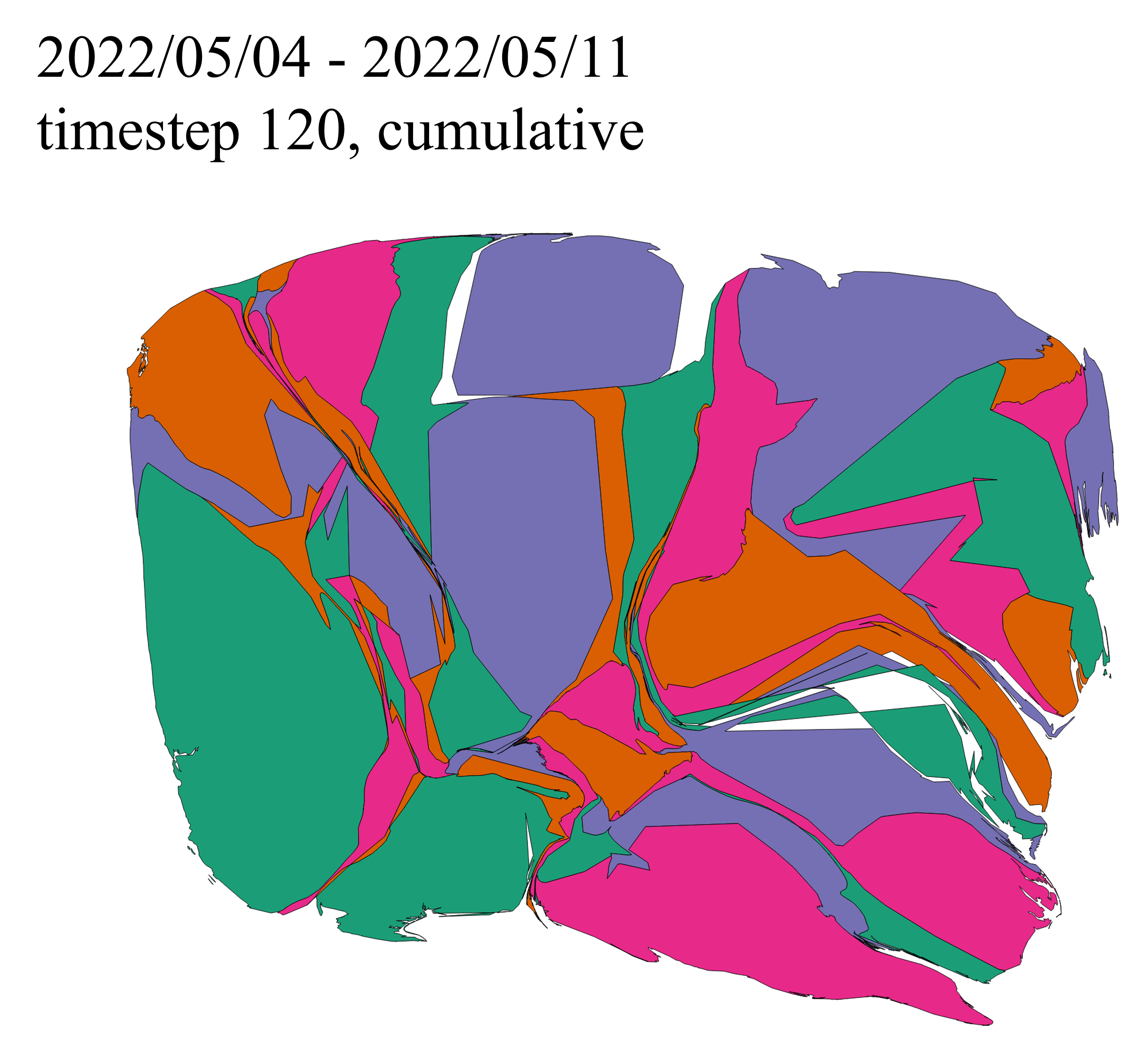}
  \caption{Cartograms showing weekly United States \textit{new} COVID-19 cases, dataset from \cite{weekly_covid_usa}. The evolution of global integral statistics $M_i(t)$ is presented at the top. The user chose the date 2020/06/10 as a starting point for analysis. For the date 2020/10/28, both direct and cumulative approaches produce similar cartograms. The cartogram computed for 2022/05/11 by the cumulative approach is overly distorted. Note the rapid changes of $M_i(t)$ shortly before 2022/05/11.}
  \label{fig:direct_vs_cumulative}
\end{figure}

Figure~\ref{fig:direct_vs_cumulative_stats} depicts the dynamics of the cartographic errors and computation time for the weekly number of \textit{total} COVID-19 cases in the United States, which exhibit gradual changes over the whole observation period. Then, the cumulative approach is beneficial as it reduces the computation time by approximately one order of magnitude when compared to the direct approach. The quality measures remain comparable or slightly higher. Thus, the choice between the two approaches depends on the variation rate of the underlying data and the desired trade-off between errors and computation time.

\renewcommand{\figsize}{0.49}
\begin{figure}[!hbt]
  \centering
  \includegraphics[width=\figsize\linewidth]{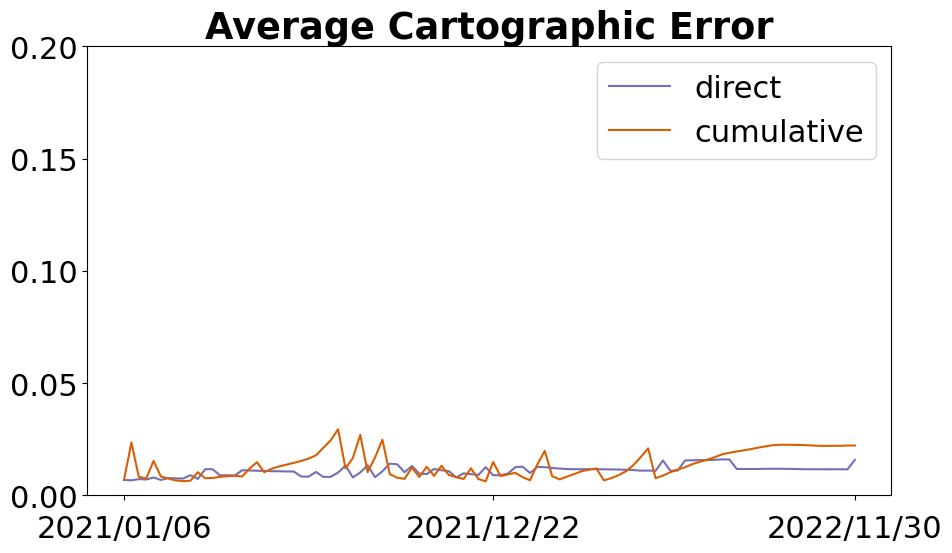}
  \includegraphics[width=\figsize\linewidth]{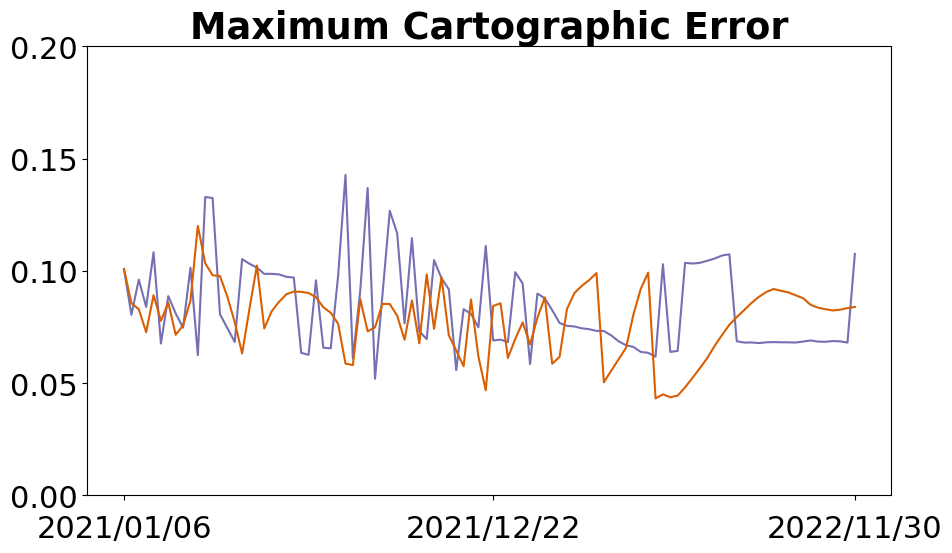}\\
  \includegraphics[width=\figsize\linewidth]{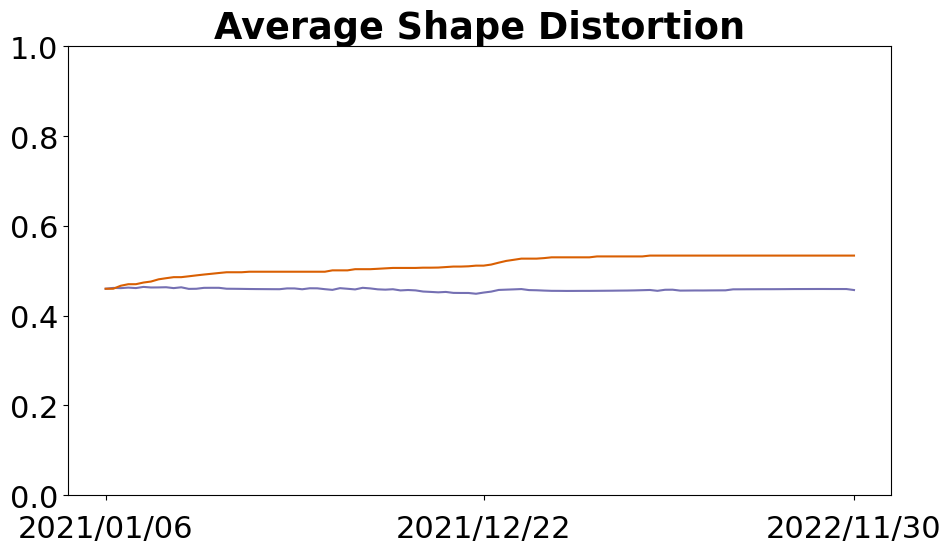}
  \includegraphics[width=\figsize\linewidth]{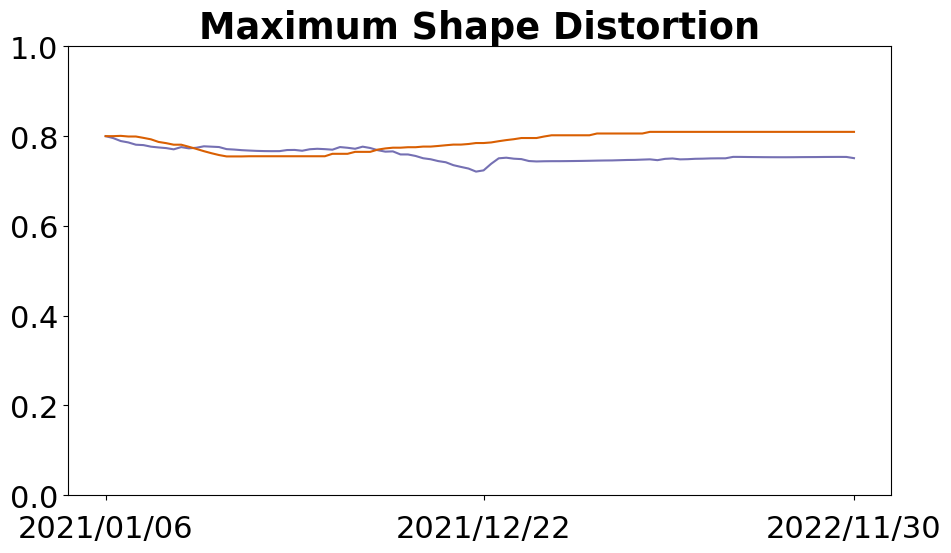}\\
  \includegraphics[width=\figsize\linewidth]{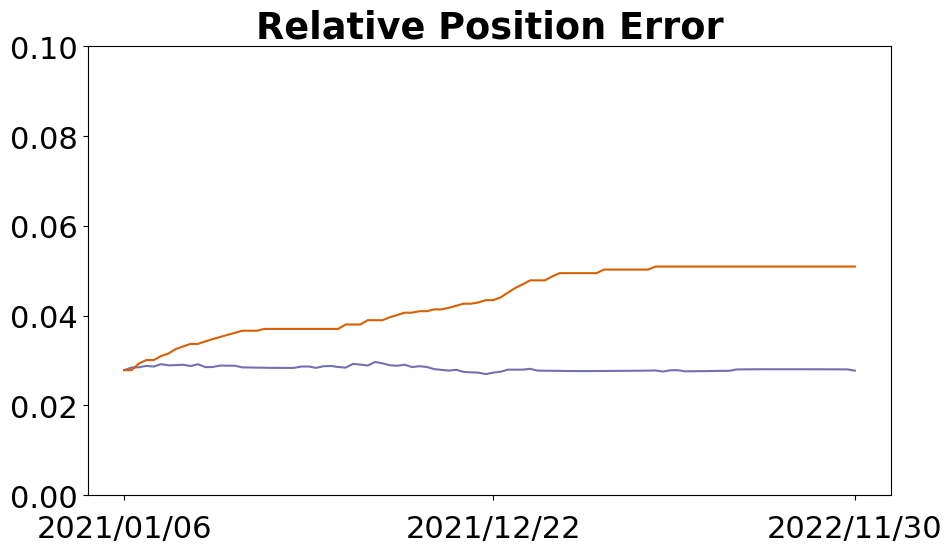}
  \includegraphics[width=\figsize\linewidth]{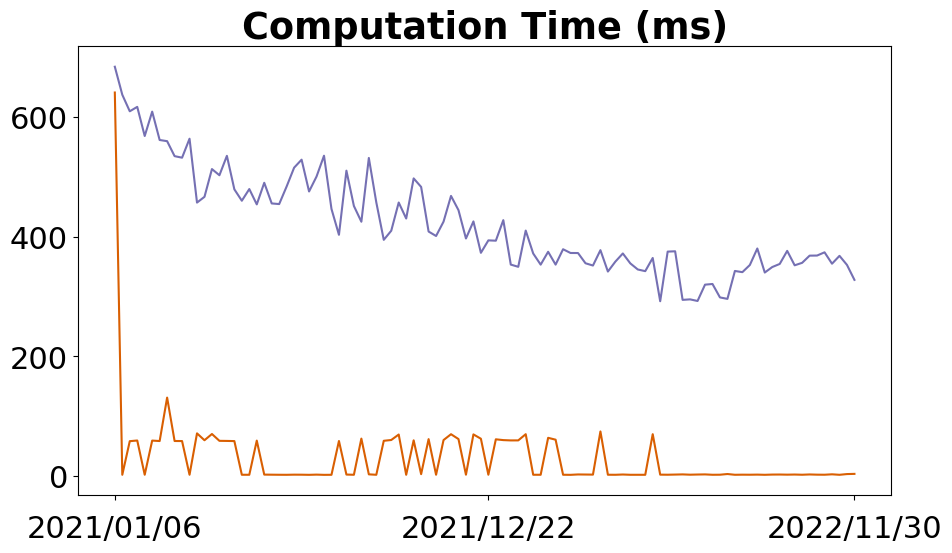}\\
  \caption{Quality measures and performance for direct and cumulative cartograms over time for
    the weekly number of \textit{total} COVID-19 cases in the United States. The cumulative approach allows for computing cartograms significantly faster than the direct approach (bottom right), while the cartographic errors remain comparable for both approaches (first row). However, we notice an increase of shape distortions (middle row) and the relative position error (bottom left).
  }
  \label{fig:direct_vs_cumulative_stats}
\end{figure}

\subsection{Dynamics of Global Integral}\label{sec:integral}

The global integral $M_i(t)$ represents the sum of the geostatistic data over all regions for time $t$. The cartograms presented in Figures~\ref{fig:dynamic_cartogram} and~\ref{fig:direct_vs_cumulative} depict the relative regional statistics at each distinct time moment but do not allow for comparing the total number of COVID-19 cases over time: The area of the map remains approximately constant while $M_i(t)$ varies. We overcome this issue by properly adjusting total background integral $M_b$(t) as presented in Section~\ref{sec:steering}. We allow the user to define the area fraction range $[A_{i,\mathrm{min}}, \,A_{i,\mathrm{max}}]$ for the cartogram map to vary. In the example presented in Figure~\ref{fig:dynamic_area}, the total German GDP is encoded in the cartogram area. Here, the area fraction range is $[0.1,\,0.9]$.

\renewcommand{\figsize}{0.485}
\begin{figure}[!hbt]
  \centering
  \includegraphics[width=0.95\linewidth]{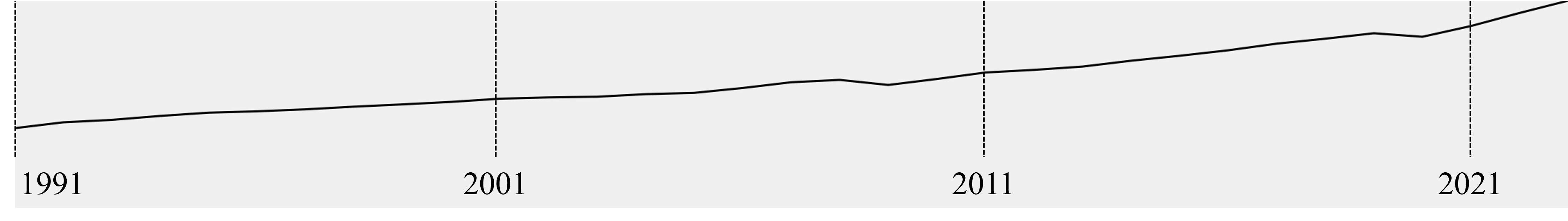}\\
  \includegraphics[width=\figsize\linewidth]{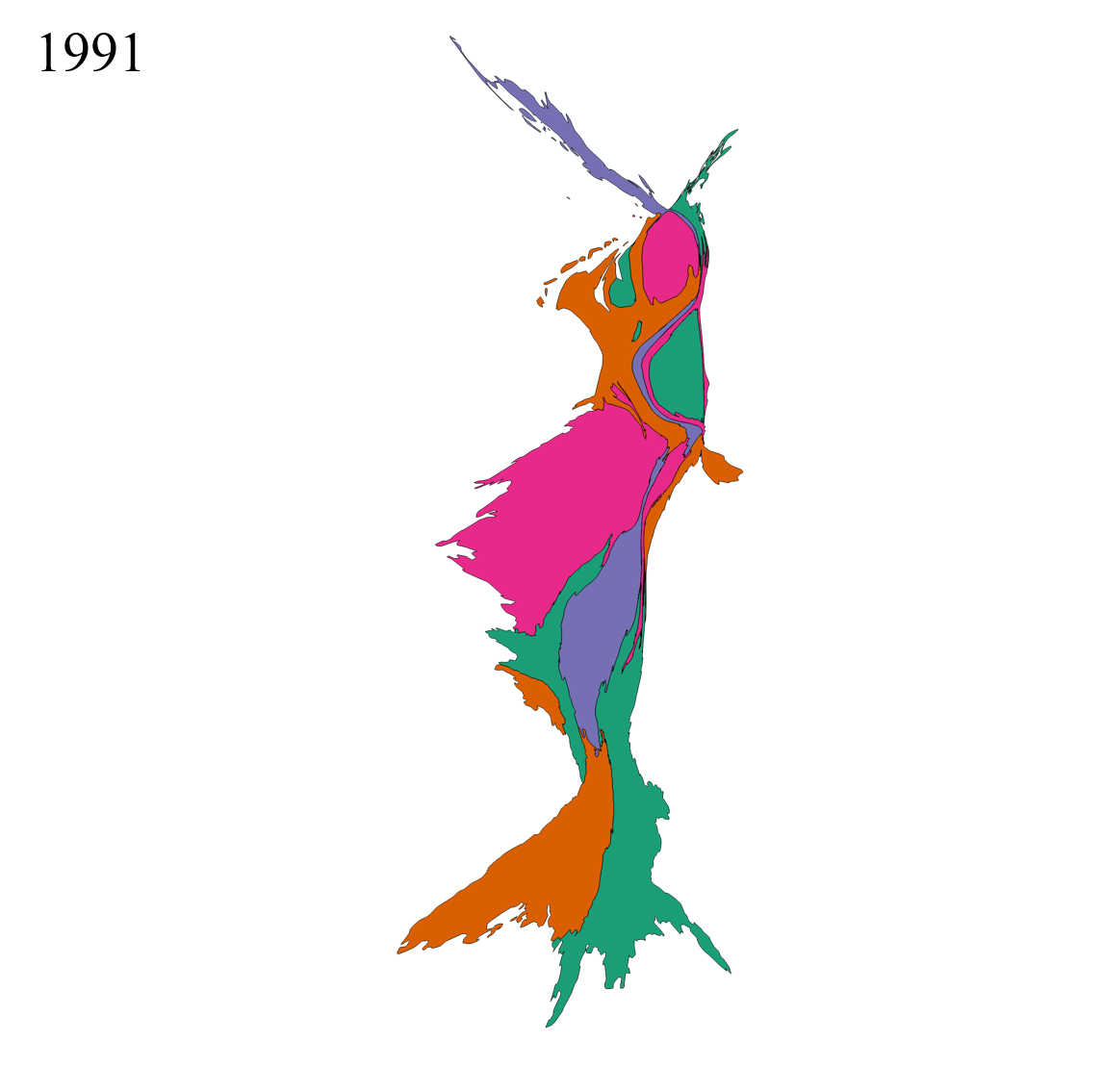}
  \includegraphics[width=\figsize\linewidth]{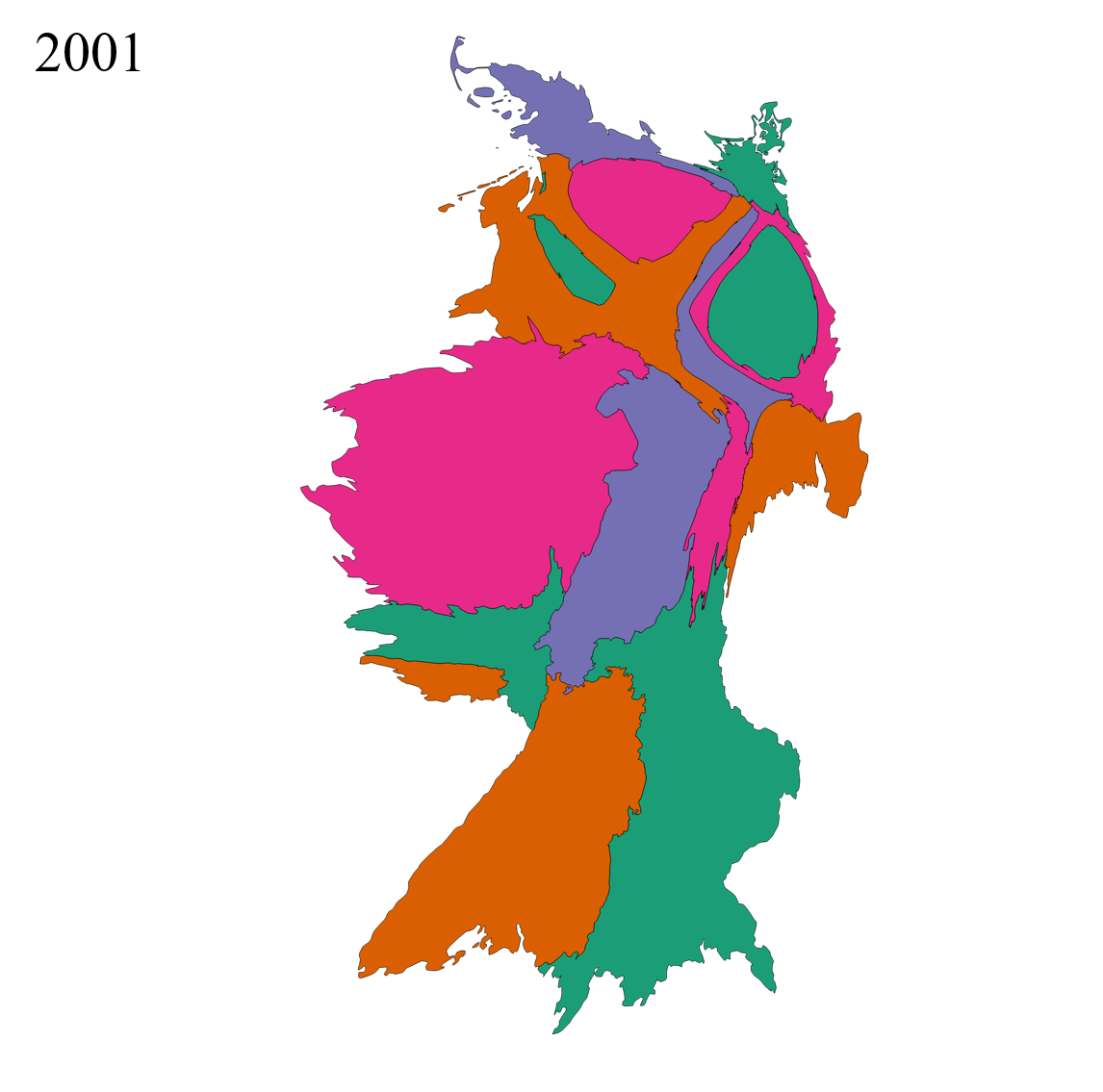}\\
  \includegraphics[width=\figsize\linewidth]{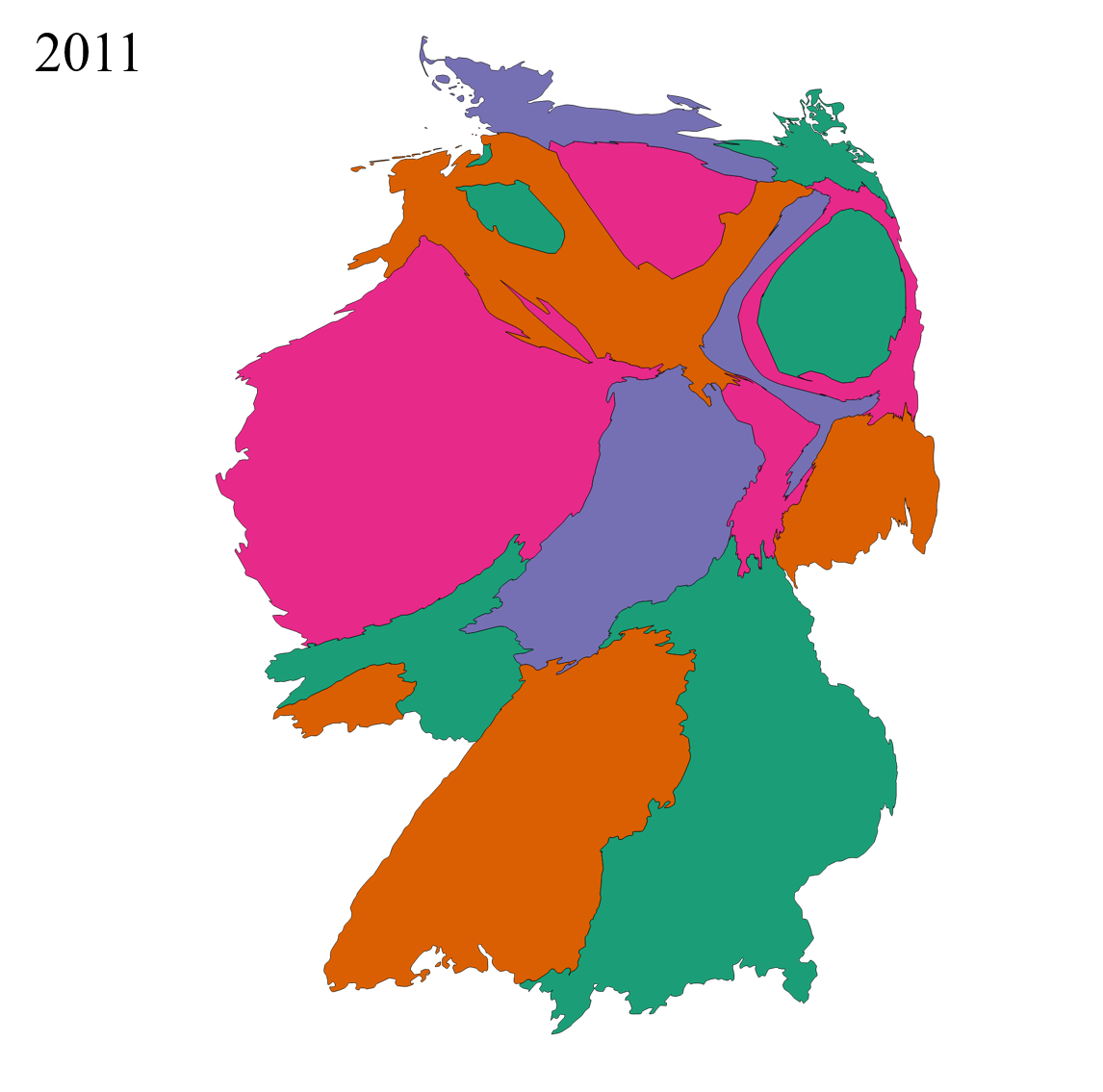}
  \includegraphics[width=\figsize\linewidth]{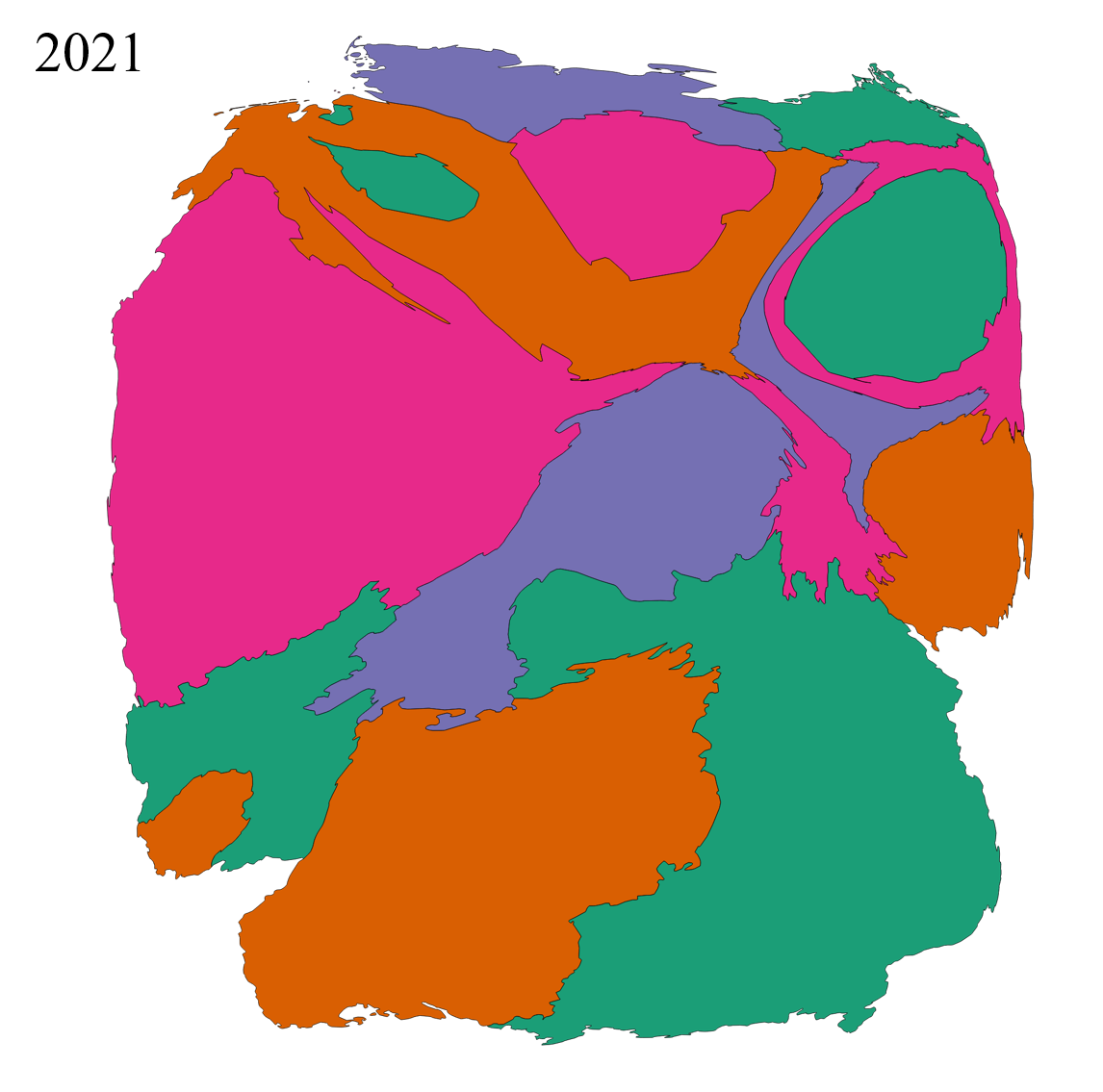}
  \caption{Yearly gross domestic product in Germany (dataset from \cite{germany_gdp}). Our directly computed cartograms show an absolute increase over time.}
  \label{fig:dynamic_area}
\end{figure}

\section{Conclusion} \label{sec:conclusion}

We presented a fast algorithm for constructing time-varying contiguous cartograms. The algorithm relies on DET~\eqref{map_res} developed by Rave et al.~\cite{Rave24_TVCG} applied to the density texture similar to the one used by Molchanov and Linsen~\cite{Molchanov20_wscg} for computing pseudo-cartograms. The density regularization is effectively performed in two dimensions, thus overcoming the separability limitation of the methods by Tobler~\cite{Tobler86} and Bak et al.~\cite{Bak09}.

While rubber-sheet methods for computing cartograms typically cannot guarantee the preservation of the original topology, physics-based algorithms may solve this issue using the continuity equation in computations. Both approaches try to ensure the smoothness of the distortion by imposing local conditions on the mapping (non-overlapping or continuity, respectively). Our method, in contrast, relies on global information about the density distribution encoded in InIms. By design, InIms are monotonic, as they accumulate the underlying non-negative density function over progressively expanding domains. The resulting spatial transformation is globally smooth. In the cartograms computed using the proposed method, we observed no violations of the original map topology. A rigorous proof of topology preservation is left for future work.

The flow-based method by Gastner et al.~\cite{Gastner18} requires adaptive time steps and repeated integrations to achieve the desired accuracy. The algorithm is not quite flexible to the variation of the background density value and a wide boundary layer is needed to mitigate the effects of the artificial periodicity of the boundary conditions, as reported by the authors. Our proposed method does not involve temporal integrations with unknown steps, demonstrates smaller or comparable cartographic errors, and is significantly faster. The calculations do not require any boundary layer, allowing for resolving the regions of interest more precisely. However, the shape of the visual domain boundary may influence the shape of the cartogram. This effect can be mitigated by carefully selecting the BDV. The user may interactively control the BDV, thus, balancing the cartographic error and shape preservation in the resulting map.

Short execution times allow for using the proposed method in visualizing temporal data and morphing cartograms constructed for different statistics in run time. We extensively addressed the issue of the cartograms' stability and coherence by proposing and comparing two approaches: direct and cumulative. The former applies DET to the original map, while the latter reuses the cartogram layout with updated geostatistic data. We demonstrated that, for moderately changing temporal data, the cumulative approach leads to a significant reduction of computation time. However, for long-time evolution data with rapid changes, geometric and topological errors may accumulate and result in dramatic distortions. We further developed a method for reflecting the dynamics of the global statistics as the total area of the cartogram and introduced interactive tools for the user to steer the visualization. Our source code will be made publicly available when the paper is published.

\section*{Acknowledgments}
\noindent This work was funded by the Deutsche Forschungsgemeinschaft (DFG) – \mbox{MO 3050/2-3} and \mbox{CRC 1450 – 431460824}.

\bibliographystyle{IEEEtran}
\bibliography{references1,references2}

\cleardoublepage
\includepdf[pages=-]{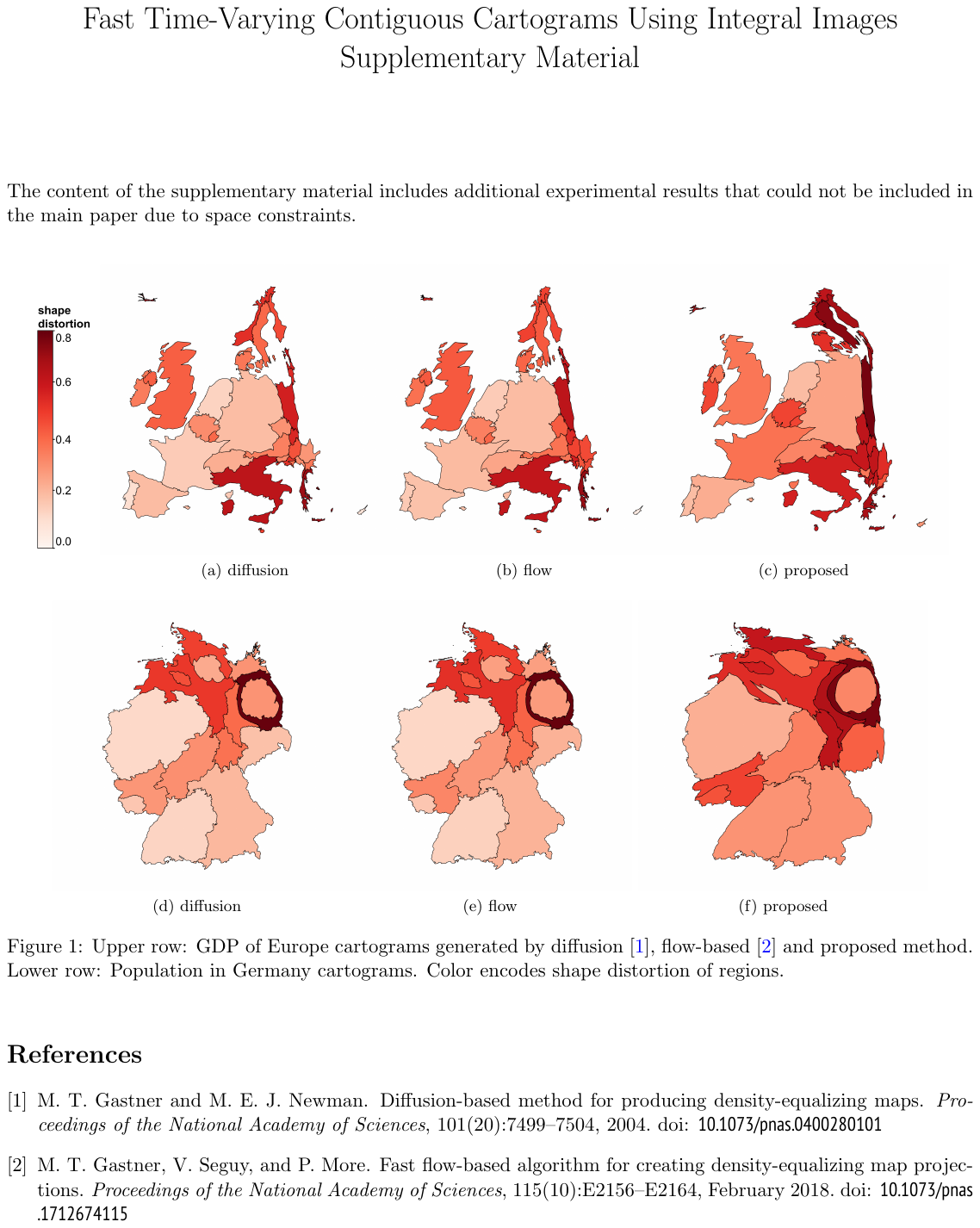}

\end{document}